%% file: main_new.tex
\documentclass[aps,pra,reprint,longbibliography,amsmath,amssymb,floatfix,nofootinbib]{revtex4-2}

\usepackage{graphicx,bm,amsthm,xcolor}
\usepackage[expansion=false]{microtype}
\usepackage[colorlinks=true,linkcolor=blue!50!black,citecolor=blue!50!black,urlcolor=blue!50!black]{hyperref}

\newtheorem{proposition}{Proposition}
\newtheorem{lemma}{Lemma}
\newtheorem{theorem}{Theorem}
\theoremstyle{remark}

\theoremstyle{plain}

\DeclareMathOperator{\Tr}{Tr}
\DeclareMathOperator{\spec}{spec}

\newcommand{\id}{\mathbb I}
\newcommand{\tp}{{\mathsf T}}
\newcommand{\D}{\mathcal D}

\newcommand{\vacS}{\Omega_S}

\newcommand{\Mslow}{\mathcal M}
\newcommand{\Ccal}{\mathcal C}
\newcommand{\nucrit}{\nu_{\mathrm{crit}}}
\hypersetup{pdftitle={Graph spectra and late-time concurrence in a lossy reservoir network},pdfauthor={Haoyuan Tian and Gourab Ghoshal}}

\begin{document}

\raggedbottom

\title{Graph spectra and late-time concurrence in a lossy reservoir network}

\author{Haoyuan Tian}
\affiliation{Department of Mathematics, University of Rochester, Rochester, New York 14627, USA}

\author{Gourab Ghoshal}
\affiliation{Department of Physics and Astronomy, University of Rochester, Rochester, New York 14627, USA}

\begin{abstract}

We investigate how coupling the local environments of noninteracting qubits into a network changes their late-time entanglement dynamics. Each qubit interacts with a lossy bosonic memory mode, with coherent hopping between memory modes determined by an underlying graph. The initial state contains at most two excitations. We show that the graph spectrum selects the collective modes that survive longest, thereby controlling both the decay and spatial distribution of pair concurrence. At resonance, nonbipartite graphs generically approach a fixed pattern of relative concurrence determined by the Perron eigenvector, whereas bipartite graphs support interference between two equally damped modes. This interference can produce arbitrarily late entanglement revivals when it overcomes the exponential decay of the surviving modes. We derive explicit thresholds separating revival and eventual monotonic decay, and show that loss from two-excitation states can drive the system across these thresholds without changing the graph or the initial one-excitation state. Underlying these results is a general bound on the sector decay rates, $\gamma_2\ge2\gamma_1$, valid at arbitrary uniform detuning. We also identify an exceptional equal-rate regime in which coherence between the vacuum and two-excitation sectors contributes to the leading concurrence. Together, these results establish how the topology of a structured dissipative environment and the redistribution of excitations through loss determine the long-time behavior of entanglement.
\end{abstract}
\maketitle

\section{Introduction}
\label{sec:introduction}

Bellomo, Lo Franco, and Compagno showed that two noninteracting qubits, each coupled to an independent non-Markovian reservoir, can recover entanglement~\cite{Bellomo2007} after a finite interval of complete disentanglement~\cite{YuEberly2004,Aolita2015}. Their construction also applies to many noninteracting qubits with independent local reservoirs. Here we ask how coupling the local memory modes~\cite{Garraway1997,Tamascelli2018} changes the late-time pair concurrence~\cite{Wootters1998} of $N$ qubits, and whether this behavior can be understood from the spectral structure of the resulting reservoir network.

Collective dynamics can arise through coupling to a shared environment or through feedback, as demonstrated in studies of chemical and other coupled oscillator systems~\cite{Ghoshal2016,GarciaSelfa2021,Sowinski2024}. Here we examine how coupling the local environments of otherwise noninteracting qubits into a network changes their entanglement dynamics. We formulate the problem using coupled bosonic modes~\cite{Nokkala2016,Pleasance2020,Medina2021,Huang2026}, motivated by cavity and circuit QED systems~\cite{SchmidtKoch2013,Kollar2019,Sundaresan2019,Ferreira2021}.
We consider a finite, simple, connected, undirected graph with \(N\ge3\) vertices, each hosting a qubit coupled to a bosonic memory mode, and the graph edges specify coherent hopping between the memory modes. Only the memory modes lose excitations, through independent local channels at zero temperature. This defines a lossy Jaynes-Cummings-Hubbard model~\cite{Greentree2006,Angelakis2007,SchmidtKoch2013} on the graph. We consider initially empty memory modes and arbitrary qubit states with at most two excitations.

Coupled cavities have been studied as auxiliary environments that prolong entanglement~\cite{Man2015,Basit2018}, as lossy lattices that mediate interactions between emitters~\cite{Roccati2022,Roccati2024}, and as arrays supporting one-excitation transport~\cite{Makin2009} and atomic entanglement distribution~\cite{Mendonca2020,Zurita2025}. Di Fidio, Ares, and Sperling obtained exact one-excitation propagation in general lossy cavity networks with one atom per cavity~\cite{DiFidio2024}. Beyond the one-excitation sector, earlier work includes two-polariton states in closed chains~\cite{Li2015,WongLaw2011}, collective decay from one or two localized excitations~\cite{Memarzadeh2013}, approximately additive decay rates of multiexcitation subradiant states in emitter chains~\cite{Albrecht2019,ZhangMolmer2019,Poshakinskiy2021}, and two-excitation bound states that outlive all one-excitation states~\cite{ZhangYuMolmer2020}.

With two excitations initially present, loss contributes to the one-excitation block of the density matrix~\cite{TanasFicek2004,Mazzola2009}. Relating late-time concurrence to the graph spectrum therefore requires accounting for this contribution and controlling the decay of the two-excitation sector. We establish that the smallest amplitude decay rates in the two sectors satisfy $\gamma_2\ge 2\gamma_1$ for arbitrary uniform detuning and finite positive coupling, and determine the conditions for equality. At nonnegative detuning the inequality is strict, allowing the leading concurrence to be expressed in terms of the least damped one-excitation modes, with coefficients that retain contributions from two-excitation loss.

The physical mechanism follows from the collective modes of the reservoir network, reflecting the broader role of graph spectra in organizing spatial patterns in networked dynamical systems~\cite{Mimar2019,Mimar2021,Stich2013}. The adjacency eigenvalues shift the frequencies of the collective memory modes relative to the qubits. Channels with larger effective detuning have least damped eigenmodes with smaller amplitudes in the lossy memory degrees of freedom, and consequently longer lifetimes. At resonance, the spectral radius therefore determines the slowest decay rate, while the Perron vector fixes the asymptotic ratios of pair concurrence on nonbipartite graphs when the slowest mode is populated. 
On bipartite graphs, spectral symmetry preserves two equally damped modes with different frequencies. Their interference can produce arbitrarily late revivals when it overcomes the exponential decay envelope. Here a revival means renewed growth of concurrence after a decrease, not necessarily following a period of complete disentanglement~\cite{Ficek2006,Das2009}. Loss from two excitations can populate these long-lived modes and change the revival behavior even though the two-excitation sector decays faster. We also identify an exceptional detuned regime where $\gamma_2=2\gamma_1$ and both sectors contribute to the leading concurrence.

\section{Model and spectral channels}
\label{sec:model}

We consider a finite simple connected undirected graph $G$ with $N\ge3$ vertices and adjacency matrix $A$. Each vertex $r\in V=\{1,\dots,N\}$ hosts a qubit with transition frequency $\omega_q$ and a bosonic memory mode with bare frequency $\omega_m$. The qubit has ground state $|g_r\rangle$ and excited state $|e_r\rangle$ with raising and lowering operators $\sigma_r^+=|e_r\rangle\langle g_r|$ and $\sigma_r^-=(\sigma_r^+)^\dagger$. The memory mode has annihilation operator $m_r$ with $[m_r,m_s^\dagger]=\delta_{rs}$ and $[m_r,m_s]=0$. In the frame rotating at $\omega_q$, with $\hbar=1$, the Jaynes-Cummings-Hubbard Hamiltonian is
\begin{equation}
\begin{aligned}
H={}&\Delta\sum_r m_r^\dagger m_r+g\sum_r\left(\sigma_r^+m_r+\sigma_r^-m_r^\dagger\right)\\
&+\eta\sum_{r,s}A_{rs}m_r^\dagger m_s,
\end{aligned}
\label{eq:H}
\end{equation}
where $\Delta=\omega_m-\omega_q$ is the uniform detuning, with resonance at $\Delta=0$, and $g,\eta>0$ are uniform coupling strengths. Here $A_{rs}=1$ if vertices $r$ and $s$ are joined by an edge, and zero otherwise.

\begin{figure*}[t]
\centering
\includegraphics[width=\textwidth]{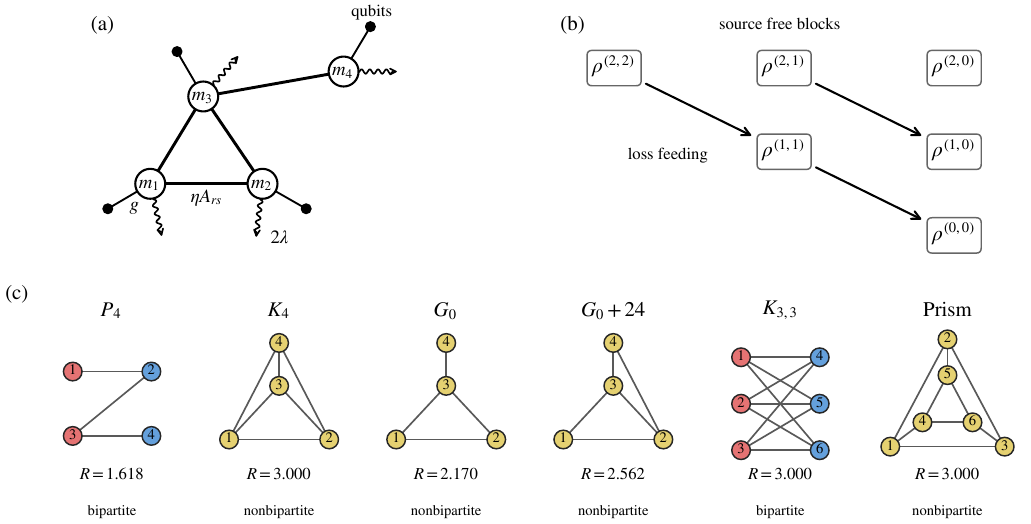}
\caption{(a) Each qubit couples to a lossy memory mode. The memory modes hop on $G_0$ and lose population at rate $2\lambda$. (b) Loss feeds lower excitation blocks. The source-free blocks in the top row receive no feeding from higher excitation sectors. Conjugate blocks are omitted. (c) Representative graphs and their spectral radii $R$. The graph $G_0$ has edges $12,13,23,34$, and $G_0+24$ also has the edge joining vertices $2$ and $4$. Red and blue vertices form the two parts of each bipartite graph. Nonbipartite graphs have yellow vertices. Vertices $1,2,3$ form one triangle of the prism.}
\label{fig:model}
\end{figure*}

Only the memory modes lose excitations, through independent local channels at zero temperature [Fig.~\ref{fig:model}(a)]. With $S$ denoting all qubits and $M$ all memory modes, the joint state $\rho_{SM}$ obeys the Lindblad equation~\cite{Lindblad1976,GoriniKossakowskiSudarshan1976}
\begin{equation}
\dot{\rho}_{SM}=-i[H,\rho_{SM}]+2\lambda\sum_r\D[m_r]\rho_{SM},
\label{eq:master}
\end{equation}
where $\lambda>0$ and $\D[L]\rho=L\rho L^\dagger-\frac{1}{2}\{L^\dagger L,\rho\}$.

The memory modes are initially in their joint vacuum $|\Omega_M\rangle$, uncorrelated with the qubits,
\begin{equation}
\rho_{SM}(0)=\rho_S(0)\otimes|\Omega_M\rangle\langle\Omega_M|,
\label{eq:initial}
\end{equation}
where $\rho_S(0)$ is an arbitrary density operator supported on $\mathcal Q_0\oplus\mathcal Q_1\oplus\mathcal Q_2$, and $\mathcal Q_n$ is the span of the qubit states with exactly $n$ excitations. With $|\Omega_S\rangle=\bigotimes_r|g_r\rangle$ the qubit vacuum, this subspace is spanned by $|\Omega_S\rangle$, $|r\rangle=\sigma_r^+|\Omega_S\rangle$ for $r\in V$, and $|rs\rangle=\sigma_r^+\sigma_s^+|\Omega_S\rangle$ for $r<s$.

The total excitation operator is
\begin{equation}
\mathcal N=\sum_r\left(\sigma_r^+\sigma_r^-+m_r^\dagger m_r\right).
\label{eq:excitation-number}
\end{equation}
Since $[H,\mathcal N]=0$ and $[\mathcal N,m_r]=-m_r$, the Hamiltonian preserves total excitation number and each loss lowers it by one.

Equation~\eqref{eq:master} is equivalent to
\begin{equation}
\begin{aligned}
\dot{\rho}_{SM}={}&-i\left(H_{\mathrm{eff}}\rho_{SM}-\rho_{SM}H_{\mathrm{eff}}^\dagger\right)\\
&+2\lambda\sum_r m_r\rho_{SM}m_r^\dagger,
\end{aligned}
\label{eq:unravel}
\end{equation}
with
\begin{equation}
H_{\mathrm{eff}}=H-i\lambda\sum_r m_r^\dagger m_r.
\label{eq:effective-Hamiltonian}
\end{equation}
The operator $H_{\mathrm{eff}}$ generates the unnormalized evolution conditioned on no quantum jumps.

Let $\Pi_n$ project onto the sector with $n$ total excitations. Since $[H_{\mathrm{eff}},\mathcal N]=0$, the conditional evolution preserves each excitation sector, with generator
\begin{equation}
H_n=\left.\Pi_n H_{\mathrm{eff}}\Pi_n\right|_{\operatorname{ran}\Pi_n}.
\label{eq:sector-generator}
\end{equation}
The smallest amplitude decay rate in each excited sector is
\begin{equation}
\gamma_n=\min_{z\in\spec H_n}\left(-\operatorname{Im}z\right),\qquad n=1,2.
\label{eq:rates}
\end{equation}
An eigenvector of $H_n$ with eigenvalue $z$ defines an $n$-excitation mode with frequency $\operatorname{Re}z$. Modes with decay rate $\gamma_n$ are called least damped.

Let $\rho^{(n,m)}=\Pi_n\rho_{SM}\Pi_m$ denote the $(n,m)$ block of the joint density matrix, and write $|0\rangle=|\Omega_S\rangle\otimes|\Omega_M\rangle$. With $H_0=0$ and $\Gamma_r^{(n)}=\sqrt{2\lambda}\Pi_{n-1}m_r\Pi_n$,
\begin{equation}
\begin{split}
\dot\rho^{(n,m)}={}&-i\bigl(H_n\rho^{(n,m)}-\rho^{(n,m)}H_m^\dagger\bigr)\\
&+\sum_r\Gamma_r^{(n+1)}\rho^{(n+1,m+1)}\Gamma_r^{(m+1)\dagger},
\end{split}
\label{eq:general-blocks}
\end{equation}
for $0\le n,m\le2$, with $\rho^{(n,m)}=0$ for $n>2$ or $m>2$. Such excitation-sector cascades are used in exact treatments of Lindblad equations without gain~\cite{Torres2014}. We refer to the sum over $r$ as loss feeding: dissipation transfers contributions from higher to lower excitation blocks [Fig.~\ref{fig:model}(b)], while the conditional evolution within each sector is governed by $H_n$.

In the ordered basis
\begin{equation}
\mathcal B_1=\bigl(\sigma_1^+|0\rangle,\dots,\sigma_N^+|0\rangle, m_1^\dagger|0\rangle,\dots,m_N^\dagger|0\rangle\bigr),
\label{eq:single-excitation-basis}
\end{equation}
the sector generator is
\begin{equation}
H_1=
\begin{pmatrix}
0 & g\id_N\\
g\id_N & \eta A+(\Delta-i\lambda)\id_N
\end{pmatrix},
\label{eq:H1}
\end{equation}
where $\id_N$ is the $N\times N$ identity matrix. The spectral decomposition of the adjacency matrix is
\begin{equation}
A=\sum_{\theta\in\spec A}\theta E_\theta,
\label{eq:adjacency-decomposition}
\end{equation}
where $E_\theta$ is the orthogonal projector onto the eigenspace with eigenvalue $\theta$.

For a normalized eigenvector $u$ of $A$ with eigenvalue $\theta$, define
\begin{equation}
\begin{aligned}
|q_u\rangle&=\sum_r u_r|r\rangle\otimes|\Omega_M\rangle,\\
|m_u\rangle&=|\Omega_S\rangle\otimes\sum_r u_r m_r^\dagger|\Omega_M\rangle.
\end{aligned}
\label{eq:channel-basis}
\end{equation}
Using $Au=\theta u$, we obtain
\begin{equation}
\begin{aligned}
H_1|q_u\rangle&=g|m_u\rangle,\\
H_1|m_u\rangle&=g|q_u\rangle+(\Delta+\eta\theta-i\lambda)|m_u\rangle.
\end{aligned}
\label{eq:channel-action}
\end{equation}
Thus $\mathcal S_u=\operatorname{span}\{|q_u\rangle,|m_u\rangle\}$ is invariant under $H_1$, and in the ordered orthonormal basis $(|q_u\rangle,|m_u\rangle)$ the restriction $H_1|_{\mathcal S_u}$ is represented by
\begin{equation}
h_\theta=
\begin{pmatrix}
0 & g\\
g & D_\theta-i\lambda
\end{pmatrix},
\qquad D_\theta=\Delta+\eta\theta.
\label{eq:channel-generator}
\end{equation}
We call $h_\theta$ a spectral channel, with effective detuning $D_\theta=(\omega_m+\eta\theta)-\omega_q$. Each adjacency eigenvalue thus shifts the frequency of a collective memory mode relative to the qubits. The channel occurs in $H_1$ with multiplicity $\operatorname{rank}E_\theta$, and its eigenvalues are the roots of $z^2-(D_\theta-i\lambda)z-g^2=0$.

For either eigenvalue $z$ of $h_\theta$, let $\chi=(1,z/g)^\tp$, so that $h_\theta\chi=z\chi$. Multiplying this equation by $\chi^\dagger$ from the left and taking imaginary parts gives
\begin{equation}
-\operatorname{Im}z=\lambda\frac{|z|^2}{g^2+|z|^2}.
\label{eq:channel-damping}
\end{equation}
The decay rate is therefore proportional to the fraction of the mode residing in the lossy memory degrees of freedom. Since the roots have product $-g^2\ne0$, both decay rates are positive, so $\gamma_1>0$. For $D_\theta\ne0$ the two roots have different moduli (Appendix~\ref{app:spectral}), and by Eq.~\eqref{eq:channel-damping} the root of smaller modulus, denoted $z_\theta$, also has the smaller decay rate.

Let $\gamma(D)$ denote the smaller amplitude decay rate of a spectral channel with effective detuning $D$. It is even in $D$ and strictly decreases with $|D|$ (Appendix~\ref{app:spectral}). Thus, increasing the magnitude of the effective detuning suppresses the memory-mode component of the channel's least damped eigenmode and prolongs its lifetime. Since $\spec H_1=\bigcup_{\theta\in\spec A}\spec h_\theta$,
\begin{equation}
\gamma_1=\min_{\theta\in\spec A}\gamma(\Delta+\eta\theta).
\label{eq:single-sector-rate}
\end{equation}
The channel $h_\theta$ has an eigenvalue $z$ satisfying $-\operatorname{Im}z=\gamma_1$ if and only if $|\Delta+\eta\theta|=\max_{\vartheta\in\spec A}|\Delta+\eta\vartheta|$. At resonance, $\Delta=0$, this condition is equivalent to $|\theta|=\max_{\vartheta\in\spec A}|\vartheta|$.

Let $R$ denote the spectral radius of $A$,
\begin{equation}
R=\max_{\theta\in\spec A}|\theta|.
\label{eq:spectral-radius}
\end{equation}
Connectedness makes $R$ a simple eigenvalue with a positive unit eigenvector $v$, called the Perron vector~\cite{BrouwerHaemers2012}. A graph is bipartite if its vertices split into two parts and every edge joins opposite parts. A connected graph is nonbipartite exactly when it contains a cycle of odd length~\cite{BrouwerHaemers2012}. For a connected graph, $-R$ is an eigenvalue exactly when the graph is bipartite. In that case, let $V_\pm$ denote the two parts and $J$ the diagonal matrix with $J_{rr}=\pm1$ for $r\in V_\pm$, so that $JAJ=-A$ and $w=Jv$ is a unit eigenvector at $-R$.

\section{Least damped channels and decay rates}
\label{subsec:slow-subspaces}

Let $\theta_-=\min_{\vartheta\in\spec A}\vartheta$ and define
\begin{equation*}
\mathcal M=\{\theta\in\spec A:\gamma(\Delta+\eta\theta)=\gamma_1\}.
\end{equation*}
We call $h_\theta$ a least damped channel when $\theta\in\mathcal M$. The ordering of $\gamma(D)$ gives
\begin{equation}
\mathcal M=
\begin{cases}
\{R\},&\Delta>-\frac{\eta}{2}(R+\theta_-),\\
\{R,\theta_-\},&\Delta=-\frac{\eta}{2}(R+\theta_-),\\
\{\theta_-\},&\Delta<-\frac{\eta}{2}(R+\theta_-).
\end{cases}
\label{eq:general-slow-labels}
\end{equation}
At $\Delta=-\eta(R+\theta_-)/2$, the two extremal channels have equal decay rates, with $D_R=-D_{\theta_-}=\eta(R-\theta_-)/2$ and $z_{\theta_-}=-\overline{z_R}$.

At resonance, $\Delta=0$, we have $\mathcal M=\{R\}$ for nonbipartite graphs and $\mathcal M=\{R,-R\}$ for bipartite graphs. In both cases, every $z_\theta$ with $\theta\in\mathcal M$ has algebraic multiplicity one in $H_1$ (Appendix~\ref{app:spectral}).

For every $\theta\in\mathcal M$, we have $D_\theta\ne0$, and $z_\theta$ is the unique eigenvalue of $h_\theta$ with $-\operatorname{Im}z_\theta=\gamma_1$. Let $P_\theta$ denote the corresponding spectral projection of $H_1$, given explicitly in Appendix~\ref{app:slow-detuning}. For fixed $g,\eta,\lambda>0$ and $\Delta\in\mathbb R$, there exists $\varepsilon>0$ such that
\begin{equation}
e^{-iH_1t}=\sum_{\theta\in\mathcal M}e^{-iz_\theta t}P_\theta
+O\bigl(e^{-(\gamma_1+\varepsilon)t}\bigr),
\label{eq:general-slow-expansion}
\end{equation}
where the remainder is measured in operator norm (Appendix~\ref{app:slow-detuning}). Thus, at long times, the one-excitation dynamics is governed by the least damped spectral channels.

\begin{proposition}[Sector decay bound and its equality case]
For the model of Sec.~\ref{sec:model}, we have
\begin{equation}
\gamma_2\ge2\gamma_1>0
\label{eq:sector-rate-bound}
\end{equation}
for all $g,\eta,\lambda>0$ and $\Delta\in\mathbb R$. Equality $\gamma_2=2\gamma_1$ holds if and only if $\Delta\le-\eta(R+\theta_-)/2$ and there exists a nonzero symmetric matrix $S$ with zero diagonal satisfying $AS=\theta_-S$. In particular, if $\Delta\ge0$ or if $\theta_-$ is a simple eigenvalue of $A$, then
\begin{equation}
\gamma_2>2\gamma_1.
\label{eq:gap}
\end{equation}
\label{prop:sector-rate-bound}
\end{proposition}

The bound establishes a hierarchy between excitation sectors. When the inequality is strict, the leading concurrence at order $e^{-2\gamma_1t}$ is determined by the slow one-excitation coherence, although its coefficient can retain contributions from loss out of the two-excitation sector. This distinction underlies the asymptotic concurrence analysis below.

The proof is given in Appendix~\ref{sector-bound-proof}. For $\Delta\ge0$, strict inequality follows from $R+\theta_-\ge0$, with equality $R+\theta_-=0$ only for bipartite graphs, whose smallest adjacency eigenvalue $\theta_-=-R$ is simple. The equality condition can nevertheless be realized at negative detuning. On the complete graph $K_N$, $\theta_-=-1$, and every symmetric matrix with zero diagonal and zero row sums satisfies $AS=-S$. Nonzero matrices of this kind exist for $N\ge4$, giving $\gamma_2=2\gamma_1$ whenever $\Delta\le-\eta(N-2)/2$. We examine the case $K_4$ in Sec.~\ref{sec:equal-rates}.

\section{Late-time concurrence}
\label{sec:concurrence}

For vertices $i<j$, the pair state is
\begin{equation}
\rho_{ij}(t)=\Tr_{M,V\setminus\{i,j\}}\rho_{SM}(t),
\label{eq:pair-state}
\end{equation}
where the trace removes all memory modes and all qubits except $i,j$. We use the ordered basis $(|11\rangle,|10\rangle,|01\rangle,|00\rangle)$, with the first label referring to qubit $i$ and $1$ denoting an excitation.

Let $Y=\sigma_y\otimes\sigma_y$, where $\sigma_y$ is the Pauli $y$ matrix. For any positive semidefinite $4\times4$ matrix $X$, define
\begin{equation}
\mathcal C(X)=\max\{0,\xi_1-\xi_2-\xi_3-\xi_4\},
\label{eq:concurrence}
\end{equation}
where $\xi_1\ge\xi_2\ge\xi_3\ge\xi_4\ge0$ are the square roots of the eigenvalues of $XYX^*Y$, and complex conjugation is taken in the stated basis. For a normalized state this is the two-qubit concurrence~\cite{Wootters1998}, and the same formula defines its homogeneous extension, $\mathcal C(aX)=a\mathcal C(X)$ for $a\ge0$. We write $C_{ij}(t)=\mathcal C(\rho_{ij}(t))$.

For every such $X$,
\begin{equation}
\bigl|\mathcal C(X)-2|X_{23}|\bigr|\le2\sqrt{X_{11}X_{44}}.
\label{eq:general-bound23}
\end{equation}
Here $X_{23}=\langle10|X|01\rangle$. The lower estimate follows from the bound obtained from the diagonal and antidiagonal entries~\cite{HashemiRafsanjani2012}, and Appendix~\ref{app:pair-bounds} proves both estimates.

The pair population $(\rho_{ij})_{11}$ comes entirely from $\rho^{(2,2)}$. When $\gamma_2>2\gamma_1$, its square root decays faster than $e^{-2\gamma_1t}$, so the doubly excited pair population cannot contribute at leading order. Equation~\eqref{eq:general-bound23} then reduces the concurrence at order $e^{-2\gamma_1t}$ to $2|(\rho_{ij})_{23}|$.

Write $B_{nm}=\rho^{(n,m)}(0)$ and $|\mathsf q_i\rangle=\sigma_i^+|0\rangle$. Let $v_a$ run through orthonormal real bases of the adjacency eigenspaces with $\theta\in\mathcal M$, and let $z_a$ be the associated root $z_\theta$. The corresponding right and left eigenvectors are
\begin{equation}
x_a=\begin{pmatrix}v_a\\z_av_a/g\end{pmatrix},
\qquad y_a^\dagger=\frac{x_a^\mathsf T}{1+z_a^2/g^2}.
\end{equation}

The bound $\gamma_2\ge2\gamma_1>\gamma_1$ ensures convergence of the integral defining the matrix $Q$,
\begin{equation}
\begin{split}
Q_{ab}={}&y_a^\dagger B_{11}y_b
+\int_0^\infty e^{i(z_a-\bar z_b)u}y_a^\dagger F_{11}(u)y_b\,du,\\
F_{11}(u)={}&\sum_r\Gamma_r^{(2)}\rho^{(2,2)}(u)\Gamma_r^{(2)\dagger}.
\end{split}
\label{eq:general-Q}
\end{equation}

The positive semidefinite matrix $Q$ describes the surviving populations and coherences of the least damped one-excitation modes. Its first term projects the initial one-excitation state onto these modes, while the integral accounts for excitation transferred into them by loss from the two-excitation sector. Thus, even when two excitations decay faster, they can influence the coefficients governing late-time concurrence. Define
\begin{equation}
\Phi_{ij}(t)=\sum_{a,b}Q_{ab}v_a(i)v_b(j)e^{-i(\omega_a-\omega_b)t},\ \omega_a=\operatorname{Re}z_a.
\label{eq:general-Phi}
\end{equation}

When $\gamma_2=2\gamma_1$, contributions from both excitation sectors can survive at the same leading order. To retain these terms, define the balanced pair matrix
\begin{equation}
\begin{aligned}
\Lambda_t&=\operatorname{diag}(e^{\gamma_1t},1,1,e^{-\gamma_1t}),\\
X_{ij}(t)&=e^{2\gamma_1t}\Lambda_t\rho_{ij}(t)\Lambda_t.
\end{aligned}
\label{eq:balanced-pair-matrix}
\end{equation}
Each entry of $X_{ij}(t)$ is a finite sum $\sum_s p_s(t)e^{st}$ with distinct complex exponents $s$ and polynomial coefficients $p_s(t)$. Define $X_{ij}^{(0)}(t)$ by retaining the terms with $\operatorname{Re}s=0$.

\begin{proposition}[Asymptotic laws for exact concurrence]
For every initial state of the form Eq.~\eqref{eq:initial} and every pair, there is $\beta>0$ such that
\begin{equation}
C_{ij}(t)=e^{-2\gamma_1 t}\mathcal F_{ij}(t)
+O(e^{-(2\gamma_1+\beta)t}),
\label{eq:general-concurrence-law}
\end{equation}
where
\begin{equation}
\mathcal F_{ij}(t)=
\begin{cases}
2|\Phi_{ij}(t)|,&\gamma_2>2\gamma_1,\\
\mathcal C(X_{ij}^{(0)}(t)),&\gamma_2=2\gamma_1.
\end{cases}
\label{eq:general-profiles}
\end{equation}
\end{proposition}
Appendix~\ref{app:general-concurrence} gives the entries of $X_{ij}^{(0)}$, proves its positivity, and establishes these asymptotic laws. We call $\mathcal F_{ij}$ the leading profile, $e^{-2\gamma_1t}\mathcal F_{ij}(t)$ the leading concurrence, and $e^{2\gamma_1t}C_{ij}(t)$ the rescaled concurrence.

A revival of $C_{ij}$ is a triple of times $t_0<t_1<t_2$ with $C_{ij}(t_1)<\min\{C_{ij}(t_0),C_{ij}(t_2)\}$. We say that $C_{ij}$ has arbitrarily late revivals if it has a revival with $t_0>T$ for every $T$.

\subsection{Perron pair ratios on nonbipartite graphs}

At resonance, Eq.~\eqref{eq:general-concurrence-law} becomes
\begin{equation}
\begin{aligned}
C_{ij}(t)&=2e^{-2\gamma_1t}|\Phi_{ij}(t)|+\mathcal R_{ij}(t),\\
\mathcal R_{ij}(t)&=O(e^{-(2\gamma_1+\beta)t}).
\end{aligned}
\label{eq:late}
\end{equation}
For a nonbipartite graph, the Perron mode is the unique least damped channel, giving $\Phi_{ij}=Q_{RR}v_iv_j$. When $Q_{RR}>0$, the concurrence of every pair therefore inherits the spatial weights of the Perron eigenvector, and
\begin{equation}
\lim_{t\to\infty}\frac{C_{ij}(t)}{C_{kl}(t)}=\frac{v_iv_j}{v_kv_l}.
\label{eq:ratio}
\end{equation}
Thus, although concurrence decays in magnitude, its relative distribution across pairs approaches a pattern determined solely by the graph.

The condition $Q_{RR}>0$ is essential. The graph with edges $12,15,25,34,35,45$ contains triangles, but the initial state $(|1\rangle-|3\rangle)/\sqrt2$ has support only on the adjacency eigenvectors $(1,\pm1,-1,\mp1,0)/2$ at eigenvalues $\pm1$, so $Q_{RR}=0$. Interference between the qubit amplitudes of these two channels causes $C_{13}$ to repeatedly vanish and revive, even at arbitrarily late times (Appendix~\ref{app:unloaded}). Thus, odd cycles alone do not exclude arbitrarily late revivals.

\subsection{Perron pair ratios on bipartite graphs}

At resonance, a bipartite graph retains two equally damped channels, associated with $R$ and $-R$. Their different frequencies produce persistent interference in the leading concurrence profile when $Q_{R,-R}\ne0$.

Define
\begin{equation}
\begin{gathered}
\Omega=2|\omega_R|,\qquad U=Q_{RR}-Q_{-R,-R},\\
F=2|Q_{R,-R}|,\qquad \nu=F/\Tr Q,
\end{gathered}
\label{eq:visibility}
\end{equation}
with $\Tr Q>0$. Since $Q$ is positive semidefinite, $0\le\nu\le1$. Taking $v_R=v$ and $v_{-R}=w=Jv$ gives
\begin{align}
|\Phi_{ij}(t)|&=v_iv_j\Tr Q[1+\nu\cos(\Omega t+\phi)],
\label{eq:profile-same}\\
|\Phi_{ij}(t)|&=v_iv_j\sqrt{U^2+F^2\sin^2(\Omega t+\phi')},
\label{eq:profile-cross}
\end{align}
for vertices in the same and opposite parts, respectively. With $\omega_R<0$ from Eq.~\eqref{sector-bound-resonance}, the phases are $\phi'=\arg Q_{R,-R}$ and $\phi=\phi'+\pi(1-J_{ii})/2$, and they do not enter when $Q_{R,-R}=0$.

The time dependence is identical for all pairs within each of three classes: those entirely in $V_+$, those entirely in $V_-$, and those connecting the two parts.
Consequently, the Perron ratio in Eq.~\eqref{eq:ratio} holds between any two pairs within the same class, provided their common leading profile remains bounded away from zero. For same-part pairs this is guaranteed by $\Tr Q>F$, while for opposite-part pairs it follows from $U\ne0$.

If $Q_{RR}Q_{-R,-R}=0$, positivity of $Q$ forces $F=0$, and the temporal factor reduces to $\Tr Q$ for every pair. In this case, Eq.~\eqref{eq:ratio} holds between arbitrary pairs, including those belonging to different classes. When the common leading profile has recurring zeros, subleading contributions can become relevant near those times, and the asymptotic expansion alone does not establish the ratio limit.

\subsection{Numerical graph comparisons}
\label{sec:numerics}

We propagate Eq.~\eqref{eq:master} exactly on the space with at most two total excitations and evaluate concurrence from the complete pair state. Unless stated otherwise, $\Delta=0$, $g=0.3\lambda$, and $\eta=3\lambda$. We use the graphs $G_0$ and $G_0+24$ of Fig.~\ref{fig:model}(c), the four-vertex path $P_4$, the complete bipartite graph $K_{3,3}$, and the triangular prism. The comparisons between graphs use
\begin{equation}
|\psi_{ij}\rangle=\sqrt{0.1}e^{0.4i}|\vacS\rangle
+\sqrt{0.8}\frac{|i\rangle+|j\rangle}{\sqrt2}
+\sqrt{0.1}e^{0.7i}|ij\rangle.
\label{eq:prep}
\end{equation}
Appendix~\ref{app:numerics} describes the numerical methods and independent checks.

\begin{figure*}[t]
\centering
\includegraphics[width=\textwidth]{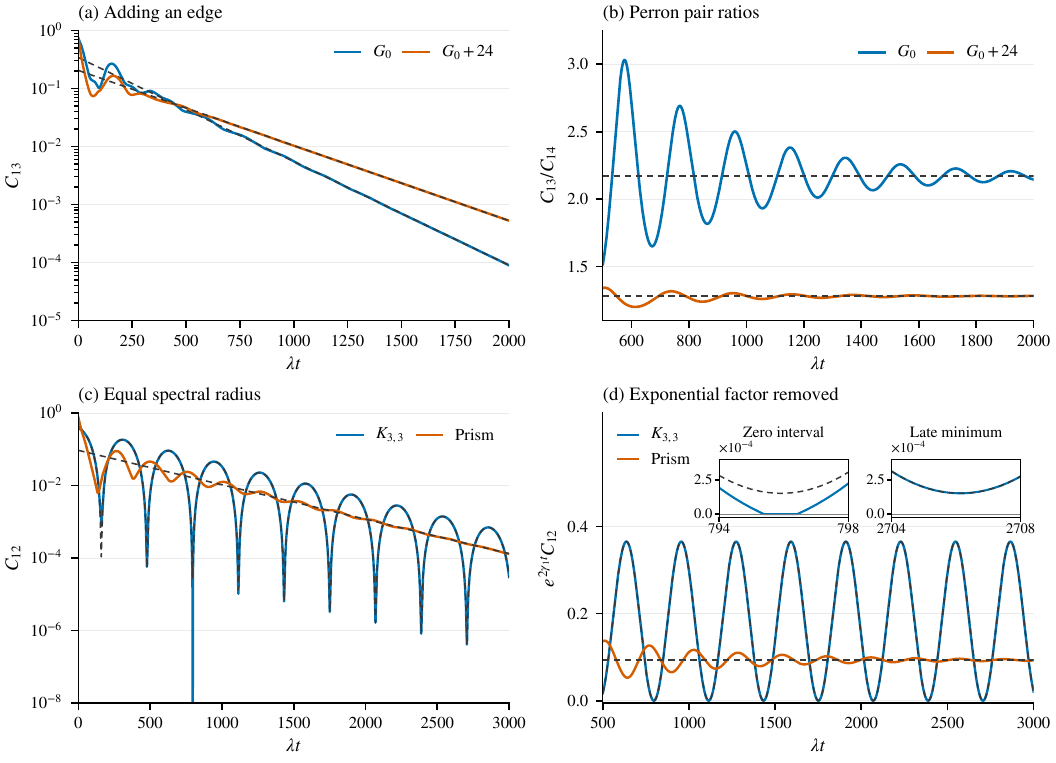}
\caption{Graph dependence of concurrence at $\Delta=0$, $g=0.3\lambda$, and $\eta=3\lambda$. Colored solid curves are numerical results, and dark dashed curves are asymptotic predictions, with the Perron limits in (b). (a,b) Preparation $|\psi_{13}\rangle$ on $G_0$ and $G_0+24$. Adding edge $24$ reduces the asymptotic decay rate, but the concurrence curves cross near $\lambda t=446$. The ratios $C_{13}/C_{14}$ approach different Perron limits. (c,d) Preparation $|\psi_{12}\rangle$ on $K_{3,3}$ and the prism. Both graphs have $R=3$ and a uniform Perron vector, but their remaining adjacency spectra differ. Their rescaled concurrences approach periodic and constant profiles, respectively. On $K_{3,3}$, $C_{12}=0$ for $795.39\le\lambda t\le796.42$. The insets in (d) share one vertical scale and show this interval and a positive late minimum.}
\label{fig:graph}
\end{figure*}

Figures~\ref{fig:graph}(a,b) show that adding edge $24$ lowers $2\gamma_1/\lambda$ from $0.004124$ to $0.002984$, but the concurrence curves nevertheless cross. Thus, a slower asymptotic decay does not guarantee greater concurrence at intermediate times. The pair ratios approach their respective Perron limits of $2.170$ and $1.281$. In Figs.~\ref{fig:graph}(c,d), the $3$-regular graphs $K_{3,3}$ and the prism share $R=3$, a uniform Perron vector, and $\gamma_1=0.001094\lambda$. Nevertheless, their long-time behavior differs: the bipartite graph retains two least damped modes and exhibits an oscillatory concurrence profile, whereas the nonbipartite graph has a single least damped mode and approaches a constant rescaled profile.

\section{Late-time monotonicity and revival thresholds}
\label{sec:revival}

\subsection{Eventual strict decrease on nonbipartite graphs}

At resonance, a nonbipartite graph has a unique least damped Perron mode. When $Q_{RR}>0$, the leading concurrence of every pair is a positive constant times $e^{-2\gamma_1t}$. In fact, the exact concurrence is eventually strictly decreasing for every pair.

Establishing strict decrease requires control not only of the remainder but also of its variation in time. For fixed $m>0$, Appendix~\ref{app:concurrence} proves that there are constants $K,T>0$ such that
\begin{equation}
|\mathcal R_{ij}(t+h)-\mathcal R_{ij}(t)|\le K e^{-(2\gamma_1+\beta)t}|h|
\label{eq:remainder}
\end{equation}
whenever $t\ge T$, $\lambda|h|\le1$, and $|\Phi_{ij}|\ge m$ on the interval between $t$ and $t+h$. When, in addition, the derivative of the leading concurrence is at most $-\delta e^{-2\gamma_1t}$ for some $\delta>0$, this bound ensures that the exact concurrence is eventually strictly decreasing. For a nonbipartite graph with $Q_{RR}>0$, $\Phi_{ij}=Q_{RR}v_iv_j$ is a positive constant, so both conditions hold.

\subsection{Revival dichotomy on bipartite graphs}

For bipartite graphs at resonance, the two least damped modes produce an oscillatory concurrence profile when $Q_{R,-R}\ne0$. Whether this produces arbitrarily late revivals depends on whether the oscillations can overcome the exponential decay envelope. Define the critical visibility
\begin{equation}
\nucrit=\frac{2\gamma_1}{\sqrt{4\gamma_1^2+\Omega^2}}.
\label{eq:nucrit}
\end{equation}
Equations~\eqref{sector-bound-resonance} and~\eqref{sector-bound-coupling} give $\nucrit=\sqrt{\gamma_1(\lambda-\gamma_1)}/g$. Since $\gamma_1=\gamma(\eta R)<\lambda/2$ and $x(\lambda-x)$ increases for $x<\lambda/2$, Eq.~\eqref{eq:channel-rate-ordering} shows that $\nucrit$ decreases strictly as $\eta R$ increases at fixed $g$ and $\lambda$.

\begin{theorem}[Revival dichotomy]
\label{thm:revival-dichotomy}
Assume that $G$ is bipartite with parts $V_+$ and $V_-$, $\Delta=0$, and $\Tr Q>0$. For a pair of vertices $i<j$, the following statements hold:
\begin{itemize}
\item If $i,j\in V_+$ or $i,j\in V_-$, then $\nu>\nucrit$ gives arbitrarily late revivals and $\nu<\nucrit$ gives eventual strict decrease.

\item Suppose $i$ and $j$ lie in opposite parts. If $U\ne0$, arbitrarily late revivals occur when
\begin{equation}
\frac{F^2}{U^2}>\frac{2\nucrit}{1-\nucrit}.
\label{eq:cross-threshold}
\end{equation}
The reverse strict inequality gives eventual strict decrease. If $U=0<F$, arbitrarily late revivals occur. If $U=F=0$, then $\mathcal F_{ij}(t)\equiv0$.
\end{itemize}
\end{theorem}

These conditions show that bipartiteness alone does not guarantee revivals: their occurrence also depends on the coherence and relative populations of the two slow modes. For example, the normalized one-excitation state $(|v\rangle+|w\rangle)/\sqrt2$, with $|v\rangle=\sum_rv_r|r\rangle$ and $|w\rangle=\sum_rw_r|r\rangle$, satisfies $\nu=1$ and $U=0<F$, so every pair has arbitrarily late revivals. 
The revival thresholds follow from comparing the maximum logarithmic growth rate of $|\Phi_{ij}|$ with the exponential decay rate $2\gamma_1$.
The dichotomy is established in Appendix~\ref{app:concurrence}, although the threshold equality cases remain unclassified. Similarly, a vanishing leading profile does not imply vanishing exact concurrence, since subleading modes may determine its late-time behavior.

\subsection{Loss feeding}

To isolate the effect of loss feeding, we fix the initial one-excitation block on $P_4$ and vary only the two-excitation and vacuum populations,
\begin{equation}
\rho_S(0)=0.15|v\rangle\langle v|+p_2|13\rangle\langle13|
+(0.85-p_2)|\vacS\rangle\langle\vacS|,
\label{eq:feeding-prep}
\end{equation}
where $0\le p_2\le0.85$ and $|v\rangle=\sum_r v_r|r\rangle$. The initial one-excitation term contributes only to $Q_{RR}$, whereas loss feeding contributes to every entry of $Q$. Increasing $p_2$ can therefore change the interference between the least damped modes without changing the initial one-excitation state.

\begin{figure*}[t]
\centering
\includegraphics[width=\textwidth]{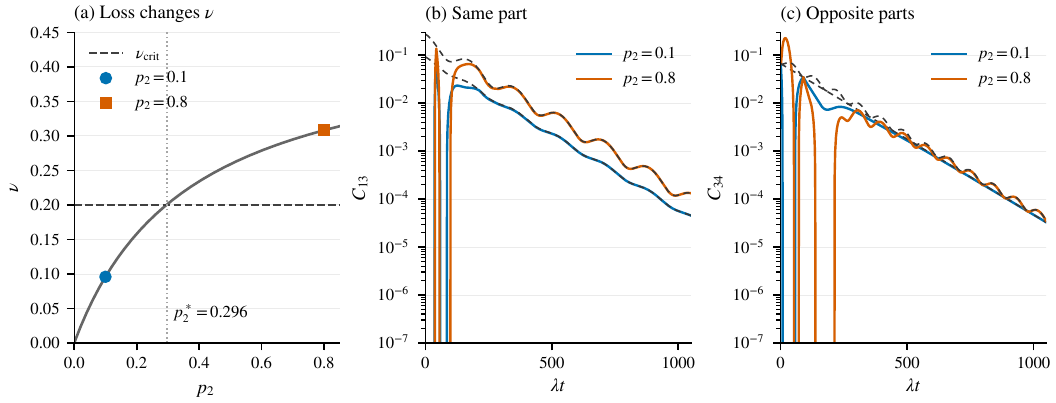}
\caption{Loss from two excitations changes both revival conditions for a fixed graph and a fixed initial one-excitation block. The state is Eq.~\eqref{eq:feeding-prep} on $P_4$, with $\Delta=0$, $g=0.3\lambda$, and $\eta=3\lambda$. (a) $\nu$ and its threshold $\nucrit$ as the initial two-excitation population $p_2$ varies. (b,c) Numerical $C_{13}$ and $C_{34}$ (solid) and asymptotic predictions (dashed) for $p_2=0.1$ and $0.8$, below and above the thresholds of Theorem~\ref{thm:revival-dichotomy}, respectively. In (b), $C_{13}=0$ for $\lambda t\le35.93$ and $58.18\le\lambda t\le84.61$ at $p_2=0.1$, and for $\lambda t\le37.42$ and $55.68\le\lambda t\le98.78$ at $p_2=0.8$. In (c), $C_{34}=0$ for $7.73\le\lambda t\le56.95$ at $p_2=0.1$, and for $53.03\le\lambda t\le71.34$ and $137.19\le\lambda t\le212.87$ at $p_2=0.8$.}
\label{fig:feeding}
\end{figure*}

For $p_2=0.1$ and $0.8$, $\nu$ takes the values $0.0959$ and $0.3082$, respectively, on opposite sides of $\nucrit=0.2004$. For the pair $34$ in opposite parts, $U$ is independent of $p_2$, while $F^2/U^2$ changes from $0.0148$ to $0.9485$, crossing the threshold $2\nucrit/(1-\nucrit)=0.5012$. The two thresholds are crossed at $p_2=0.2963$ and $0.5815$, respectively. Figure~\ref{fig:feeding} shows the corresponding changes in the exact concurrence. Without the loss-feeding integral in Eq.~\eqref{eq:general-Q}, neither transition would be predicted.

\subsection{Late revival times}
\label{sec:times}

For a bipartite graph at resonance, consider a pair in the same part with $\Tr Q>0$ and $\nucrit<\nu<1$. The leading concurrence has local maxima at
\begin{equation}
\begin{aligned}
t_n&=\frac{\vartheta-\chi-\phi+2\pi n}{\Omega},\\
\vartheta&=\arctan\frac{\Omega}{2\gamma_1},\qquad
\chi=\arccos\left(-\frac{\nucrit}{\nu}\right).
\end{aligned}
\label{eq:times}
\end{equation}
For a pair in opposite parts with $F^2(1-\nucrit)>2\nucrit U^2$, the leading concurrence has local maxima at
\begin{equation}
\begin{aligned}
t_n'&=\frac{\vartheta+\chi'-2\phi'+2\pi n}{2\Omega},\\
\chi'&=\arccos\left[\nucrit\left(1+\frac{2U^2}{F^2}\right)\right].
\end{aligned}
\label{eq:times-cross}
\end{equation}
This condition is equivalent to Eq.~\eqref{eq:cross-threshold} when $U\ne0$ and holds whenever $U=0<F$. For every sufficiently large integer $n$, the exact concurrence has at least one local maximum at $t_n+O(e^{-\beta t_n})$ for a pair in the same part and at $t_n'+O(e^{-\beta t_n'})$ for a pair in opposite parts. Appendix~\ref{app:concurrence} proves these location bounds using Eq.~\eqref{eq:remainder}.

\begin{figure*}[t]
\centering
\includegraphics[width=\textwidth]{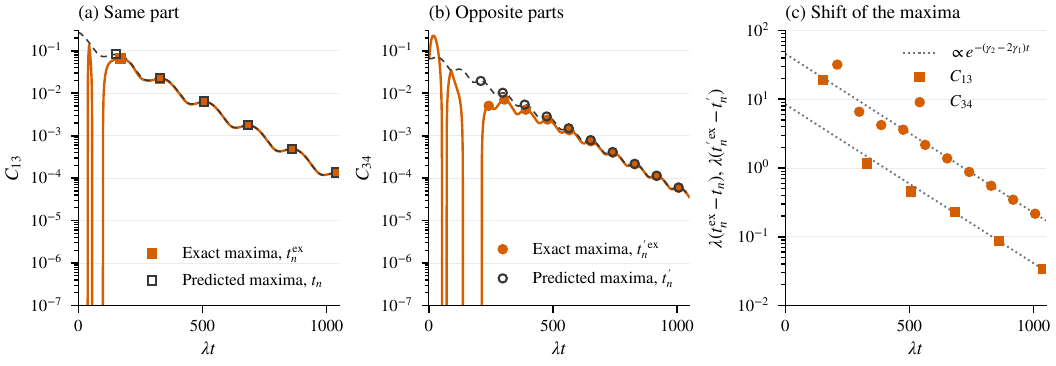}
\caption{Late revival times for Eq.~\eqref{eq:feeding-prep} on $P_4$ with $p_2=0.8$, $\Delta=0$, $g=0.3\lambda$, and $\eta=3\lambda$. (a,b) Numerical $C_{13}$ and $C_{34}$ (solid) and asymptotic predictions (dashed), as in Fig.~\ref{fig:feeding}(b,c). Filled markers show the exact local maxima at $t_n^{\mathrm{ex}}$ and $t_n^{\prime\,\mathrm{ex}}$, and open markers show the leading concurrence at $t_n$ and $t_n'$ from Eqs.~\eqref{eq:times} and \eqref{eq:times-cross}. Markers are shown for $\lambda t\ge150$. In (b), $C_{34}=0$ at the first open marker. (c) Shifts of the exact maxima from the predicted times. Dotted reference lines are proportional to $e^{-(\gamma_2-2\gamma_1)t}$ and pass through the last marker of each set, so only their slope is fixed.}
\label{fig:times}
\end{figure*}

For the preparation in Eq.~\eqref{eq:feeding-prep} with $p_2=0.8$, pair $13$ satisfies $\nucrit<\nu<1$, while pair $34$ satisfies $F^2(1-\nucrit)>2\nucrit U^2$. Figure~\ref{fig:times} compares the predicted times from Eqs.~\eqref{eq:times} and \eqref{eq:times-cross} with the local maxima $t_n^{\mathrm{ex}}$ and $t_n^{\prime\,\mathrm{ex}}$ of the exact concurrence. The first predicted maximum of $C_{34}$ shown in Fig.~\ref{fig:times} falls within the interval $137.19\le\lambda t\le212.87$, where the exact concurrence vanishes. For $150\le\lambda t\le3000$, the numerical maxima correspond one to one with the predicted times.

The computed shifts $t_n^{\mathrm{ex}}-t_n$ and $t_n^{\prime\,\mathrm{ex}}-t_n'$ are positive and exhibit approximately geometric decay. Their successive ratios are consistent with $0.3894$ for $C_{13}$ and $0.6240$ for $C_{34}$, obtained from $e^{-2\pi(\gamma_2-2\gamma_1)/\Omega}$ and $e^{-\pi(\gamma_2-2\gamma_1)/\Omega}$, respectively, where $\gamma_2-2\gamma_1=0.005324\lambda$.

\subsection{Odd cycles and detuning}

At resonance, adding an edge $ij$ gives $R'\ge v^{\mathsf T}A'v=R+2v_iv_j>R$ for the original unit Perron vector, where primes refer to the modified graph. Since $\gamma(D)$ decreases with $|D|$, the slowest decay rate satisfies $\gamma_1'<\gamma_1$ at every finite coupling. Adding an edge within one part of a bipartite graph creates an odd cycle and leaves a single least damped channel, whereas an edge joining opposite parts preserves bipartiteness and the two least damped channels.

Let $Q'$ and $C_{ij}'$ denote the matrix in Eq.~\eqref{eq:general-Q} and the concurrence for the modified graph with the same initial state. If the modified graph is nonbipartite and $Q'_{R'R'}>0$, its concurrence has a positive leading coefficient and decays more slowly than the concurrence on the original graph. 
Consequently, at sufficiently late times, the modified graph has greater concurrence than the original graph for every qubit pair, $C_{ij}'(t)>C_{ij}(t)$.

Uniform detuning provides another way to remove the degeneracy of the least damped channels without changing the graph. By Proposition~\ref{sector-bound-bipartite-detuning} of Appendix~\ref{sector-bound-proof}, a bipartite graph has $\gamma_2>2\gamma_1$ at every uniform detuning. At nonzero detuning, let $a$ label the unique least damped one-excitation mode. If $Q_{aa}>0$, the leading profile is a positive constant, and the concurrence is eventually strictly decreasing (Appendix~\ref{app:asymptotic-laws}).

For $\Delta>0$, write $\gamma_+=\gamma(\eta R+\Delta)$ and $\gamma_-=\gamma(\eta R-\Delta)$ for the rates of the channels labeled by $R$ and $-R$. Their difference satisfies
\begin{equation}
\delta\gamma=\gamma_{-}-\gamma_{+}=2\left|\frac{d\gamma}{dD}(\eta R)\right|\Delta+O(\Delta^3)
\label{eq:detuning-rate-split}
\end{equation}
as $\Delta\to0^+$. For sufficiently small detuning, $\gamma_-<2\gamma_+\le\gamma_2$, so the integrals in Eq.~\eqref{eq:general-Q}, evaluated at $z_R$ and $z_{-R}$, converge. The cross term between the two channels decays as $e^{-(\gamma_++\gamma_-)t}$, compared with $e^{-2\gamma_+t}$ for the population of the less damped channel. Its relative contribution therefore decays as $e^{-\delta\gamma t}$, suppressing the interference responsible for persistent oscillations at resonance.

\begin{figure*}[t]
\includegraphics[width=\textwidth]{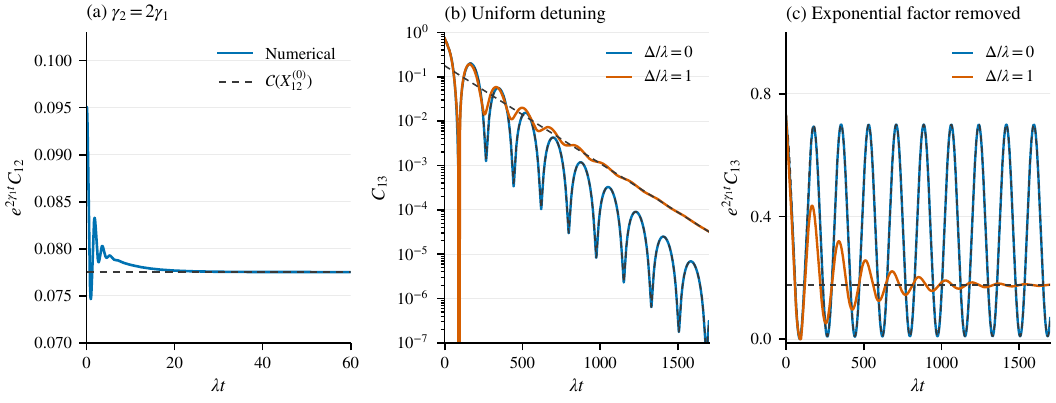}
\caption{Distinct effects of uniform detuning. Solid curves are numerical results and dashed curves are asymptotic predictions. (a) On $K_4$, $g=\eta=\lambda$ and $\Delta=-2\lambda$ give $\gamma_2=2\gamma_1$. The rescaled $C_{12}$ for Eq.~\eqref{eq:critical-preparation} with $p=0.01$ approaches the coefficient in Eq.~\eqref{eq:critical-analytic-profile}, which includes coherence between the vacuum and two qubit excitations. (b,c) Original and rescaled $C_{13}$ on $P_4$ for $\Delta=0$ and $\lambda$, with $g=0.3\lambda$, $\eta=3\lambda$, and preparation $|\psi_{13}\rangle$. Each curve in (c) uses its own $\gamma_1$. The resonant curve approaches a periodic profile, while the detuned curve approaches a constant. At resonance, $C_{13}$ remains positive, and its minimum $C_{13}=4.9\times10^{-8}$ near $\lambda t=1683$ lies below the range of (b). At $\Delta=\lambda$, $C_{13}=0$ for $87.91\le\lambda t\le96.43$. Detuning separates the amplitude decay rates of the $R$ and $-R$ channels, while finite-time revivals remain in (b).}
\label{fig:general-regimes}
\end{figure*}

Figure~\ref{fig:general-regimes}(b,c) compares $P_4$ at $\Delta=0$ and $\Delta=\lambda$, with $g=0.3\lambda$, $\eta=3\lambda$, and preparation $|\psi_{13}\rangle$. At $\Delta=\lambda$, $\mathcal M=\{(1+\sqrt5)/2\}$ and $Q_{RR}>0$, so the concurrence is eventually strictly decreasing, and the rescaled concurrence approaches $0.17630054$. Revivals nevertheless occur at finite times; after the last local maximum at $\lambda t=822.11$, the computed curve decreases up to $\lambda t=1700$.

\subsection{Equal decay rates with degenerate least damped modes}
\label{sec:equal-rates}

By Proposition~\ref{prop:sector-rate-bound}, equality $\gamma_2=2\gamma_1$ requires $\Delta\le-\eta(R+\theta_-)/2$ and a repeated eigenvalue $\theta_-$, excluding bipartite graphs. In this case, contributions from both excitation sectors can survive at the same leading order, and $\Phi_{ij}$ alone need not determine concurrence.

For the complete graph $K_4$ in Fig.~\ref{fig:model}, setting $g=\eta=\lambda=1$ and $\Delta=-2$ yields
\begin{equation}
\gamma_2=2\gamma_1=1-\sqrt{3\sqrt5-6}.
\label{eq:K4-critical-rate}
\end{equation}
Here $\mathcal M=\{-1\}$ and $\operatorname{rank}E_{-1}=3$. The eigenvalue of the least damped two-excitation modes is $2z$, where $z$ is the root of $z^2+(3+i)z-1=0$ of smaller modulus. Its algebraic and geometric multiplicities are both two. Appendix~\ref{app:critical-certificate} proves these statements using explicit eigenvectors and Lemmas~\ref{sector-bound-equality} and~\ref{sector-bound-semisimple}.

In this example, the three least damped one-excitation modes have frequency $\omega=\operatorname{Re}z$, and the two least damped two-excitation modes have frequency $2\omega$. Consequently, $X_{ij}^{(0)}(t)$ evolves by local phase rotations, leaving its concurrence constant for every initial state (Appendix~\ref{app:asymptotic-laws}).

Consider the qubit preparation
\begin{equation}
\begin{aligned}
|s\rangle&=\tfrac12(|12\rangle-|13\rangle-|24\rangle+|34\rangle),\\
|\psi_p\rangle&=\sqrt{1-p}|\Omega_S\rangle+\sqrt p|s\rangle,
\qquad 0<p<1.
\end{aligned}
\label{eq:critical-preparation}
\end{equation}
Put $\alpha=(1+z^2)^{-1}$ and $a_0=|\alpha|^2/2$. For the pair $12$, the matrix $X_{12}^{(0)}$ in Eq.~\eqref{eq:general-critical-matrix} has only diagonal and corner entries. Appendix~\ref{app:critical-preparation} evaluates these entries and gives, for $0<p<1/2$,
\begin{equation}
C_{12}(t)=2a_0[\sqrt{p(1-p)}-p]e^{-2\gamma_1t}
+O(e^{-(2\gamma_1+\beta)t}).
\label{eq:critical-analytic-profile}
\end{equation}
The coefficient is positive, and the exact concurrence is eventually strictly decreasing (Appendix~\ref{app:critical-preparation}). The leading contribution arises from coherence between the vacuum and two-qubit excitation sectors, together with populations generated by loss; the coherence between $|10\rangle$ and $|01\rangle$ vanishes.

The same preparation also illustrates why a vanishing leading profile does not determine whether exact concurrence survives. 
At $p=1/2$, the leading profile vanishes, but the exact concurrence is $|f(t)|^4/4$, where $f(t)$ denotes the qubit amplitude evolving from an initially excited qubit in the $D=-3$ channel.
It remains positive at late times and is asymptotic to $|\alpha|^4e^{-4\gamma_1t}/4$. For $p>1/2$, concurrence is instead exactly zero at all sufficiently late times. Appendix~\ref{app:critical-preparation} derives both results from the full time-dependent pair state.

Figure~\ref{fig:general-regimes}(a) tests Eq.~\eqref{eq:critical-analytic-profile} at $p=0.01$, where the rescaled concurrence approaches $\mathcal C(X_{12}^{(0)})=0.07752961$, computed from Eq.~\eqref{eq:general-critical-matrix} without fitting the trajectory.

\section{Discussion and conclusions}
\label{sec:discussion}

We have shown that coupling the local environments of noninteracting qubits into a network changes not only the rate at which entanglement decays, but also its spatial distribution and temporal behavior. The underlying mechanism is the formation of collective reservoir modes whose frequencies and decay rates depend on the graph spectrum. As the magnitude of the effective detuning increases, the least damped modes have less weight in the lossy memory degrees of freedom and consequently decay more slowly. Network structure therefore determines which collective excitations survive longest, while their populations and mutual coherence determine the entanglement that remains at late times. 

At resonance, this mechanism leads to a distinction between bipartite and nonbipartite networks. On nonbipartite graphs, the Perron mode is the unique least damped channel, and when populated, its spatial structure fixes the asymptotic ratios of pair concurrence. The concurrence of every pair then becomes strictly decreasing at sufficiently late times. Bipartite graphs instead retain two equally damped channels with different frequencies. Their interference produces an oscillatory leading concurrence profile, but arbitrarily late revivals occur when these oscillations overcome the exponential decay envelope. The resulting thresholds distinguish arbitrarily late revivals from eventual monotonic decay. Thus, networks with identical spectral radii, and hence identical slowest decay rates, can exhibit qualitatively different entanglement dynamics.

The role of dissipation becomes particularly important when the initial state contains two excitations. Although the two-excitation sector decays faster than the long-lived one-excitation modes at nonnegative detuning, loss transfers population into these modes and changes their relative weights and coherence. Consequently, changing the initial two-excitation population can move the system between revival and monotonic-decay regimes without changing the network or the initial one-excitation state. This effect is missed if the dynamics is restricted to the one-excitation sector. More generally, our bound on the sector decay rates establishes when the leading concurrence is determined by one-excitation coherence and identifies the exceptional negative-detuning regime in which both sectors contribute at the same asymptotic order.

The model also suggests possible experimental tests in coupled-cavity and circuit-QED networks, where local qubit--resonator interactions, coherent hopping, and dissipation can be engineered. Small networks with different connectivity could be used to compare the spatial distribution and temporal evolution of pair concurrence. A particularly direct test would be to vary the initial two-excitation population while keeping the network and initial one-excitation block fixed, and examine the resulting change in revival behavior. Observing these effects would require resolving concurrence before additional decoherence processes suppress the surviving collective modes. The robustness of the predicted signatures to qubit relaxation, dephasing, and parameter disorder remains to be established.

Our analysis is exact for finite connected graphs with uniform couplings, local zero-temperature losses, and initial states containing at most two excitations. These assumptions isolate the effects of collective reservoir structure from other sources of decoherence. In particular, arbitrarily late revivals do not imply sustained entanglement, since their amplitudes decrease exponentially. What persists is the interference between the least damped collective modes, whose existence and spatial structure are determined by the network. This distinction provides a basis for understanding how reservoir connectivity influences entanglement dynamics, independently of direct interactions between the qubits.

\appendix

\section{Spectral channels}
\label{app:spectral}

\subsection{Ordering of channel decay rates}

The roots $z,z'$ of $z^2-(D-i\lambda)z-g^2=0$ have product $zz'=-g^2$. If their moduli are equal, then $|z|=|z'|=g$ and $z'=-\overline z$, which implies $D=\operatorname{Re}(z+z')=0$. Thus, for $D\ne0$, there is a unique root of smaller modulus, denoted by $z$, and $|z|<g$. The root equation gives
\begin{equation}
D=\operatorname{Re}z\left(1-\frac{g^2}{|z|^2}\right).
\label{eq:root-real-part}
\end{equation}
Hence $\operatorname{Re}z<0$ for $D>0$.

For $D\ne0$ the roots are simple, and differentiating the root equation gives
\begin{equation}
\frac{dz}{dD}=\frac{z}{2z-D+i\lambda},
\label{eq:root-derivative}
\end{equation}
and therefore
\begin{equation}
\frac{d|z|^2}{dD}
=\frac{2|z|^2(2\operatorname{Re}z-D)}{|2z-D+i\lambda|^2}<0,
\qquad D>0.
\label{eq:root-modulus-ordering}
\end{equation}
Since the right-hand side of Eq.~\eqref{eq:channel-damping} is strictly increasing in $|z|^2$, the smaller decay rate $\gamma(D)$ strictly decreases for $D>0$.

The transformation $D\mapsto-D$ maps each root to $-\overline z$ and preserves its modulus and decay rate, so $\gamma(-D)=\gamma(D)$. The minimum decay rate is continuous at $D=0$. Thus
\begin{equation}
\gamma(D_1)>\gamma(D_2),
\qquad 0\le D_1<D_2.
\label{eq:channel-rate-ordering}
\end{equation}

\subsection{Least damped modes at arbitrary detuning}
\label{app:slow-detuning}

Since $R=\max_{\vartheta\in\spec A}\vartheta$ and $\eta>0$, every $\theta\in\spec A$ satisfies $D_{\theta_-}\le D_\theta\le D_R$, and $|D_\theta|<\max\{|D_{\theta_-}|,|D_R|\}$ for $\theta_-<\theta<R$. Since $\gamma(D)$ strictly decreases with $|D|$, the eigenvalues in $\mathcal M$ are determined by
\begin{equation}
D_R^2-D_{\theta_-}^2
=\eta(R-\theta_-)\bigl[2\Delta+\eta(R+\theta_-)\bigr].
\label{eq:endpoint-comparison}
\end{equation}
The factor $\eta(R-\theta_-)$ is positive, which yields Eq.~\eqref{eq:general-slow-labels}. At $\Delta=-\eta(R+\theta_-)/2$, we have $D_R=-D_{\theta_-}$, so the root symmetry gives $z_{\theta_-}=-\overline{z_R}$.

For $\theta\in\mathcal M$,
\begin{equation}
|D_\theta|=\max\{|D_{\theta_-}|,|D_R|\}
\ge\frac{\eta}{2}(R-\theta_-)>0.
\label{eq:selected-detuning-nonzero}
\end{equation}
Thus $h_\theta$ has distinct eigenvalues, $|z_\theta|<g$, and $\gamma_1<\lambda/2$. Every eigenvalue $z$ of $h_\theta$ satisfies
\begin{equation}
D_\theta=z-\frac{g^2}{z}+i\lambda,
\label{eq:detuning-from-root}
\end{equation}
so channels with different $\theta$ cannot share an eigenvalue. An orthonormal basis of $\operatorname{ran}E_\theta$ gives $\operatorname{rank}E_\theta$ mutually orthogonal invariant subspaces $\mathcal S_u$, each with matrix $h_\theta$. Hence each $z_\theta$ with $\theta\in\mathcal M$ has algebraic and geometric multiplicities $\operatorname{rank}E_\theta$ in $H_1$.

For $\theta\in\mathcal M$, let $\chi=(1,z_\theta/g)^\tp$. Since $h_\theta^\tp=h_\theta$, its spectral projection at $z_\theta$ is $\chi\chi^\tp/(\chi^\tp\chi)$, whose denominator is nonzero because $|z_\theta|<g$. Combining this projection with $E_\theta$ gives
\begin{equation}
P_\theta=\frac{1}{g^2+z_\theta^2}
\begin{pmatrix}
g^2E_\theta&gz_\theta E_\theta\\
gz_\theta E_\theta&z_\theta^2E_\theta
\end{pmatrix},
\label{eq:general-slow-projector}
\end{equation}
with $\operatorname{rank}P_\theta=\operatorname{rank}E_\theta$.

A repeated channel root requires $D_\theta=0$ and $2g=\lambda$. Then $(h_\theta+i\lambda\id_2/2)^2=0$ and
\begin{equation}
e^{-ih_\theta t}=e^{-\lambda t/2}
\left[\id_2-it\left(h_\theta+\frac{i\lambda}{2}\id_2\right)\right].
\label{eq:repeated-channel-propagator}
\end{equation}
Its decay rate satisfies $\lambda/2>\gamma_1$. All eigenvalues omitted from Eq.~\eqref{eq:general-slow-expansion} have decay rates greater than $\gamma_1$, so for fixed parameters, choosing $\varepsilon>0$ smaller than every such rate difference absorbs the possible factor $t$ and proves the expansion. The same bound holds for the first time derivative of the remainder.

At $\Delta=0$ we have $\mathcal M\subseteq\{R,-R\}$, and connectedness makes $R$ simple. When $-R\in\spec A$, the graph is bipartite, and $JAJ=-A$ makes $-R$ simple~\cite{BrouwerHaemers2012}. The multiplicity result above therefore gives algebraic multiplicity one for every $z_\theta$ with $\theta\in\mathcal M$.

\subsection{An exact equal-rate case}
\label{app:critical-certificate}

For $K_4$ with $g=\eta=\lambda=1$ and $\Delta=-2$, the adjacency eigenvalues are $3$ and $-1$, the latter with effective detuning $-3$. Let $S$ be a symmetric matrix with zero diagonal and $S\bm 1=0$, where $\bm 1$ has every entry equal to one. Such matrices form a two-dimensional space on $K_4$. If $z$ is the root of $z^2+(3+i)z-1=0$ of smaller modulus, then
\begin{equation}
|\Psi_S\rangle=\sum_{r<s}S_{rs}
(\sigma_r^++z m_r^\dagger)
(\sigma_s^++z m_s^\dagger)|0\rangle
\label{eq:K4-critical-vector}
\end{equation}
is an eigenvector of $H_2$ with eigenvalue $2z$. Together with Eq.~\eqref{eq:sector-rate-bound}, this proves Eq.~\eqref{eq:K4-critical-rate}. The map $S\mapsto|\Psi_S\rangle$ is injective through its two-qubit component and therefore gives two independent eigenvectors.

Only $h_{-1}$ is a least damped channel, so every least damped one-excitation mode has eigenvalue $z$. By the proof of Lemma~\ref{sector-bound-equality}, every eigenvector of $H_2$ with decay rate $2\gamma_1$ is an eigenvector of the auxiliary generator $\widetilde H_2$ used in that proof, with eigenvalue $2z$ and zero qubit double-occupation amplitudes. Both channels have $D_\theta\ne0$, so $H_1$ is diagonalizable, and such a vector is a symmetric combination of products of the least damped modes $(u,zu)$ with $u\perp\bm 1$. Its two-qubit component is therefore a symmetric matrix $S$ with zero diagonal and $S\bm 1=0$, and the vector equals $|\Psi_S\rangle$. Hence $2z$ has geometric multiplicity two, and Lemma~\ref{sector-bound-semisimple} gives algebraic multiplicity two.

\subsection{A preparation with \texorpdfstring{$Q_{RR}=0$}{Q(R,R) = 0}}
\label{app:unloaded}

For the five-vertex graph and initial state given in Sec.~\ref{sec:concurrence}, the state lies in the span of the two adjacency eigenvectors $u_{\pm1}=(1,\pm1,-1,\mp1,0)/2$. Let $z,z'$ be the channel roots at $D=\eta>0$, with $|z|<|z'|$. Starting from a qubit excitation, this channel has qubit amplitude
\begin{equation}
f(t)=\frac{z'e^{-izt}-ze^{-iz't}}{z'-z},
\end{equation}
and the channel at $D=-\eta$ has qubit amplitude $\overline{f(t)}$ by the root symmetry in Appendix~\ref{app:spectral}. The qubit amplitudes therefore satisfy $c_1=-c_3=\operatorname{Re}f(t)/\sqrt2$, so $C_{13}(t)=[\operatorname{Re}f(t)]^2$. Since the second term in $f$ decays faster and $\operatorname{Re}z\ne0$, this real part changes sign infinitely often at isolated zeros, which gives arbitrarily late revivals on a nonbipartite graph.

\subsection{Exact evolution of the equal-rate preparation}
\label{app:critical-preparation}

For the preparation in Eq.~\eqref{eq:critical-preparation}, let $S$ be the symmetric matrix with nonzero upper-triangular entries $(S_{12},S_{13},S_{24},S_{34})=(1,-1,-1,1)$. Then $|s\rangle=\sum_{k<l}S_{kl}|kl\rangle/2$ is normalized. Define $f(t),h(t)$ by
\begin{equation}
\begin{pmatrix}f(t)\\h(t)\end{pmatrix}
=e^{-ih_{-1}t}\begin{pmatrix}1\\0\end{pmatrix}.
\end{equation}
The zero diagonal and zero row sums of $S$ give the exact conditional evolution
\begin{equation}
\begin{aligned}
e^{-iH_2t}|s\rangle={}&\frac12\sum_{k<l}S_{kl}[f(t)\sigma_k^++h(t)m_k^\dagger]\\
&\times[f(t)\sigma_l^++h(t)m_l^\dagger]|0\rangle.
\end{aligned}
\label{eq:critical-product-evolution}
\end{equation}
A loss at vertex $r$ and time $u$ leaves the one-excitation vector whose spatial amplitudes form row $r$ of $S$ and whose channel amplitudes are $f(u),h(u)$, multiplied by $\sqrt{2\lambda}h(u)/2$. Since every row of $S$ belongs to the $-1$ adjacency eigenspace, subsequent propagation gives channel amplitudes $f(t),h(t)$. The single-channel norm satisfies
\begin{equation}
\frac{d}{dt}(|f|^2+|h|^2)=-2\lambda|h|^2,
\ 2\lambda\int_0^\infty|h(u)|^2\,du=1.
\end{equation}

The contribution to $f$ from the eigenvalue $z$ is $\alpha e^{-izt}$, with $\alpha=(1+z^2)^{-1}$. The qubit block of $e^{2\gamma_1t}\rho^{(1,1)}(t)$ therefore converges to $p|\alpha|^2S^{\mathsf T}S/4$. For the pair $12$, the identities $(S^{\mathsf T}S)_{11}=(S^{\mathsf T}S)_{22}=2$ and $(S^{\mathsf T}S)_{12}=0$ give $m_{11}=m_{22}=pa_0$ and $m_{12}=0$. Equation~\eqref{eq:critical-product-evolution} yields $d=p|\alpha|^4/4=pa_0^2$ and $b=\sqrt{p(1-p)}\alpha^2e^{-2i\omega t}/2$.

If the only nonzero off-diagonal entries of a positive semidefinite $X$ are $X_{14}$ and $X_{41}$, Eq.~\eqref{eq:concurrence} gives $\mathcal C(X)=2\max\{0,|X_{14}|-\sqrt{X_{22}X_{33}}\}$. Hence $\mathcal C(X_{12}^{(0)})=2\max\{0,|b|-pa_0\}$, which proves Eq.~\eqref{eq:critical-analytic-profile}.

The same construction gives the pair state at every time. Write $I(t)=2\lambda\int_0^t|h(u)|^2du=1-|f(t)|^2-|h(t)|^2$ and suppress the time argument of $f$ and $h$. The pair population $(\rho_{12})_{22}$ receives the contributions $p(|f|^4/4+|f|^2|h|^2/2)$ without loss and $p|f|^2I/2$ after one loss. The same holds for entry $33$. Orthogonality of the first two rows of $S$ makes entry $23$ zero. Thus
\begin{equation}
\begin{aligned}
(\rho_{12})_{11}&=p|f|^4/4,\\
(\rho_{12})_{22}=(\rho_{12})_{33}&=p(|f|^2/2-|f|^4/4),\\
(\rho_{12})_{14}&=\sqrt{p(1-p)}f^2/2,\\
(\rho_{12})_{44}&=1-p|f|^2+p|f|^4/4.
\end{aligned}
\end{equation}
Since the preparation has no one-excitation component, Eq.~\eqref{eq:general-blocks} keeps the $(2,1)$ and $(1,0)$ blocks zero, so only the diagonal and corner entries are nonzero. The exact concurrence is
\begin{equation}
C_{12}(t)=\left[|f|^2\bigl(\sqrt{p(1-p)}-p\bigr)+\frac p2|f|^4\right]_+,
\label{eq:critical-exact-time}
\end{equation}
where $[x]_+=\max(0,x)$. For $0<p<1/2$, the coefficient of $|f|^2$ is positive. Since $f-\alpha e^{-izt}=(1-\alpha)e^{-iz't}$, where $z'$ is the other root of the channel and $\operatorname{Im}z'=-(\lambda-\gamma_1)$, we have $d|f|^2/dt=-2\gamma_1|\alpha|^2e^{-2\gamma_1t}+O(e^{-\lambda t})$. As $2\gamma_1<\lambda$, $|f|^2$ and hence $C_{12}(t)$ are eventually strictly decreasing. At $p=1/2$, this is $|f|^4/4\sim|\alpha|^4e^{-4\gamma_1t}/4$. For $p>1/2$, the coefficient of $|f|^2$ is negative and $f\to0$, so the expression is zero for every sufficiently late time.

\input{sector_rate_bound}

\section{Proofs for late-time concurrence}
\label{app:general-concurrence}

We prove the asymptotic laws for $\gamma_2>2\gamma_1$ and $\gamma_2=2\gamma_1$.

We use $\|B\|_1=\operatorname{Tr}\sqrt{B^\dagger B}$ for the trace norm and $\|B\|_{\rm HS}^2=\operatorname{Tr}(B^\dagger B)$ for the Hilbert-Schmidt norm, and $\|B\|$ is the operator norm.

\subsection{Bounds for a general pair matrix}
\label{app:pair-bounds}

Factor $X=LL^\dagger$ with $L$ a $4\times4$ matrix, and denote its rows by $u_1,\dots,u_4$. The Wootters roots $\xi_k$ in Eq.~\eqref{eq:concurrence} are the singular values of
\begin{equation*}
\begin{aligned}
W=L^\tp YL={}&u_2^\tp u_3+u_3^\tp u_2-u_1^\tp u_4-u_4^\tp u_1.
\end{aligned}
\end{equation*}
Let $W_0=u_2^\tp u_3+u_3^\tp u_2$ and denote its ordered singular values by $s_k$. Its two possibly nonzero singular values are $\|u_2\|\|u_3\|\pm|u_2u_3^\dagger|$, so $s_1-s_2=2|X_{23}|$ and $s_3=s_4=0$. The singular value perturbation inequality~\cite{HornJohnson2013} gives
\begin{equation*}
\begin{aligned}
\bigl|\mathcal C(X)-2|X_{23}|\bigr|
&\le\sum_{k=1}^4|\xi_k-s_k|\\
&\le\|W-W_0\|_1\\
&\le2\|u_1\|\|u_4\|\\
&=2\sqrt{X_{11}X_{44}},
\end{aligned}
\end{equation*}
where the row norms are Euclidean. This proves Eq.~\eqref{eq:general-bound23}.

\subsection{Leading pair matrix at equal rates}
\label{app:equal-profile}

Let $P_n$ be the spectral projection onto the generalized eigenspaces of $H_n$ with decay rate $\gamma_n$. Its kernel is the sum of the remaining generalized eigenspaces. In particular, $P_1=\sum_{\theta\in\mathcal M}P_\theta$. Define
\begin{equation}
T_n(t)=e^{\gamma_nt}e^{-iH_nt}P_n.
\label{eq:general-slow-propagators}
\end{equation}
For $\theta\in\mathcal M$, $z_\theta$ is a semisimple eigenvalue of $H_1$. At $\gamma_2=2\gamma_1$, Lemma~\ref{sector-bound-semisimple} gives the same for the eigenvalues of $H_2$ with decay rate $2\gamma_1$. By Eq.~\eqref{eq:general-slow-labels}, and since $z_{\theta_-}=-\overline{z_R}$ when both channels are least damped, the roots $z_\theta$ with $\theta\in\mathcal M$ have real parts in $\{\omega,-\omega\}$, where $\omega=\operatorname{Re}z_\theta$ for any $\theta\in\mathcal M$. By the proof of Lemma~\ref{sector-bound-equality}, every eigenvector of $H_2$ with decay rate $2\gamma_1$ is an eigenvector of $\widetilde H_2$, so its eigenvalue is a sum of two such roots. Thus $T_1$, and $T_2$ at $\gamma_2=2\gamma_1$, are finite sums of constant matrices times $e^{ik\omega t}$ with integer $k$. Consequently $\Phi_{ij}$, the entries of $X_{ij}^{(0)}$ given below, and hence every leading profile $\mathcal F_{ij}$ are periodic with period $2\pi/|\omega|$. At $\gamma_2=2\gamma_1$, which requires $\Delta\ne0$, set
\begin{equation}
K_{11}=\sum_{a,b}x_aQ_{ab}x_b^\dagger,
\label{eq:general-K11}
\end{equation}
and define
\begin{equation}
\begin{split}
K_{10}&=P_1B_{10}+\int_0^\infty e^{iH_1u}P_1F_{10}(u)\,du,\\
F_{10}(u)&=\sum_r\Gamma_r^{(2)}\rho^{(2,1)}(u)\Gamma_r^{(1)\dagger},
\end{split}
\label{eq:general-K10}
\end{equation}
whose convergence is proved below. In the matrix elements below, $|ij\rangle=\sigma_i^+\sigma_j^+|0\rangle$. Suppress the time arguments of $T_1,T_2$ and put
\begin{align}
d&=\langle ij|T_2B_{22}T_2^\dagger|ij\rangle,\nonumber\\
b&=\langle ij|T_2B_{20}|0\rangle,\nonumber\\
t_k&=\langle ij|T_2B_{21}T_1^\dagger|\mathsf q_k\rangle,\nonumber\\
m_{kl}&=\langle\mathsf q_k|T_1K_{11}T_1^\dagger|\mathsf q_l\rangle,\nonumber\\
a_k&=\langle\mathsf q_k|T_1K_{10}|0\rangle,\qquad k,l\in\{i,j\}.
\end{align}
The matrix $X_{ij}^{(0)}$ defined in Sec.~\ref{sec:concurrence} has the form
\begin{equation}
X_{ij}^{(0)}(t)=
\begin{pmatrix}
d&t_i&t_j&b\\
\bar t_i&m_{ii}&m_{ij}&a_i\\
\bar t_j&\bar m_{ij}&m_{jj}&a_j\\
\bar b&\bar a_i&\bar a_j&1
\end{pmatrix}.
\label{eq:general-critical-matrix}
\end{equation}
The proof below shows that this matrix is positive semidefinite for every $t$. Feeding from $\rho^{(2,1)}$ contributes to $a_i,a_j$, and these entries cannot be discarded for general initial states.

\subsection{Asymptotic laws}
\label{app:asymptotic-laws}

\paragraph{Convergence of the integrals.}
The operator in Eq.~\eqref{eq:general-K11} equals $P_1B_{11}P_1^\dagger$ plus the integral of $e^{iH_1u}P_1F_{11}(u)P_1^\dagger e^{-iH_1^\dagger u}$. The integrand is positive semidefinite and bounded by a polynomial times $e^{-2(\gamma_2-\gamma_1)u}$, which is integrable by Eq.~\eqref{eq:sector-rate-bound}. Thus $Q\succeq0$ and the integral in Eq.~\eqref{eq:general-Q} converges. Since $\rho^{(2,1)}(u)=e^{-iH_2u}B_{21}e^{iH_1^\dagger u}$, the integrand defining $K_{10}$ is bounded by a polynomial times $e^{-\gamma_2u}$, and its integral also converges.

\paragraph{Positivity and the asymptotic law at equal rates.}
Factor $\rho_{SM}(0)=LL^\dagger$ and put $L_n=\Pi_nL$. With $\mathcal U_n(u)=e^{-iH_nu}$, form the columns
\begin{equation}
V_0=\begin{pmatrix}P_2L_2\\P_1L_1\\L_0\end{pmatrix},
\
V_r(u)=\begin{pmatrix}
0\\e^{iH_1u}P_1\Gamma_r^{(2)}\mathcal U_2(u)L_2\\
\Gamma_r^{(1)}\mathcal U_1(u)L_1
\end{pmatrix}.
\end{equation}
The Gram sum
\begin{equation}
Z=V_0V_0^\dagger+\sum_r\int_0^\infty V_r(u)V_r(u)^\dagger\,du
+\operatorname{diag}(0,0,\Tr B_{22})
\label{eq:general-critical-Gram}
\end{equation}
is positive semidefinite. Its middle row contains $K_{11}$ on the diagonal and $K_{10}$ in the vacuum column. Its vacuum entry is $\Tr B_{00}+\Tr B_{11}+\Tr B_{22}=1$, where the integral of the vacuum entry from $V_r$ equals $\Tr B_{11}$ because the initial one-excitation population is eventually lost. The other blocks are the corresponding projected initial blocks. Conjugating $Z$ by $\operatorname{diag}(T_2,T_1,1)$ and restricting to $(|ij\rangle,|\mathsf q_i\rangle,|\mathsf q_j\rangle,|0\rangle)$ gives exactly the matrix in Eq.~\eqref{eq:general-critical-matrix}, which is therefore positive semidefinite.

Write $\gamma=\gamma_1$ and use $\Lambda_t$ and $X_{ij}$ from Eq.~\eqref{eq:balanced-pair-matrix}. Since $\Lambda_tY\Lambda_t=Y$, the local filtering identity~\cite{Verstraete2001} gives $C_{ij}=e^{-2\gamma t}\mathcal C(X_{ij})$ exactly. The blocks $(2,2)$, $(2,1)$, and $(2,0)$ have no feeding term in Eq.~\eqref{eq:general-blocks}, and their projections by $P_1$ and $P_2$, scaled as in Eq.~\eqref{eq:balanced-pair-matrix}, give $d$, $t_k$, and $b$. Variation of constants for the $(1,1)$ and $(1,0)$ blocks yields $m_{kl}$ and $a_k$. The pair ground population tends to one. On this scale, every traced external excitation adds a strictly decaying factor, and the remaining spectral terms and the tails of the convergent integrals defining $Q$ and $K_{10}$ also decay exponentially. Thus, for some $\delta>0$,
\begin{equation}
\|X_{ij}-X_{ij}^{(0)}\|_1
+\|\dot X_{ij}-\dot X_{ij}^{(0)}\|_1
\le Ke^{-\delta t}.
\label{eq:general-critical-error}
\end{equation}
Polynomial factors from Jordan blocks or from resonant feeding integrals are absorbed by a strict reduction of $\delta$. For positive semidefinite matrices $A,B$, square-root factors and the singular value trace norm inequality give
\begin{equation}
|\mathcal C(A)-\mathcal C(B)|
\le(\sqrt{\Tr A}+\sqrt{\Tr B})\sqrt{\|A-B\|_1}.
\label{eq:general-Holder}
\end{equation}
We used $\|\sqrt A-\sqrt B\|_{\rm HS}^2\le\|A-B\|_1$~\cite{PowersStormer1970}. Since the traces of $X_{ij}$ and $X_{ij}^{(0)}$ are bounded, Eqs.~\eqref{eq:general-critical-error} and \eqref{eq:general-Holder} prove Eq.~\eqref{eq:general-concurrence-law} at $\gamma_2=2\gamma_1$.

\paragraph{The strict inequality.}
For $\gamma_2>2\gamma_1$, choose $2\gamma_1<\gamma_2'<\gamma_2$. Projecting with $P_1$ in the variation of constants formula for the $(1,1)$ block gives $(\rho_{ij})_{23}=e^{-2\gamma_1t}[\Phi_{ij}+O(e^{-\delta t})]$. The pair entry $11$ belongs entirely to $\rho^{(2,2)}$ and is $O(e^{-2\gamma_2't})$, so Eq.~\eqref{eq:general-bound23} proves the first branch.

\paragraph{Revival criteria with a controlled remainder.}
Let $s_0<s_1<s_2$ be fixed nonnegative times. Suppose $\tau_n\to\infty$ and, for $a=0,2$,
\begin{equation}
e^{-2\gamma_1 s_a}\mathcal F_{ij}(\tau_n+s_a)
-e^{-2\gamma_1 s_1}\mathcal F_{ij}(\tau_n+s_1)
\ge\delta_n>0,
\label{eq:general-witness}
\end{equation}
where $e^{\beta\tau_n}\delta_n\to\infty$. Equation~\eqref{eq:general-concurrence-law} then gives arbitrarily late revivals of the exact concurrence. Since every $\mathcal F_{ij}$ is periodic, it suffices that both differences in Eq.~\eqref{eq:general-witness} are positive at $\tau_n=0$, because taking $\tau_n$ as multiples of the period then yields a constant $\delta_n$.

\paragraph{Strict decrease with one least damped mode.}

Assume $\gamma_2>2\gamma_1$ and $Q_{bb}=0$ for all $b\ne a$. The rows of $L_\star$ indexed by $|10\rangle$ and $|01\rangle$ are linearly dependent. The argument in Appendix D proves Eq. (38) for any $\Delta\in\mathbb R$. Here $\Phi_{ij}=Q_{aa}v_a(i)v_a(j)$ is constant. If $\Phi_{ij}\ne0$, then $C_{ij}(t)$ is eventually strictly decreasing.

\paragraph{A constant profile with multiple least damped modes at equal rates.}
Suppose all least damped one-excitation modes have frequency $\omega$ and all least damped two-excitation modes have frequency $2\omega$, with arbitrary multiplicities. Then
\begin{equation}
X_{ij}^{(0)}(t)=U_\omega(t)X_{ij}^{(0)}(0)U_\omega(t)^\dagger,
\end{equation}
where $U_\omega=\operatorname{diag}(e^{-2i\omega t},e^{-i\omega t},e^{-i\omega t},1)$ is a product of local phase rotations. Thus $\mathcal F_{ij}$ is constant for every initial state.

\section{Variation bounds at resonance}
\label{app:concurrence}
\label{app:law}
\label{app:proof}
Assume $\Delta=0$ and write $\gamma=\gamma_1$. All constants below refer to fixed parameters and initial state, and $K$ may differ between occurrences. By Eq.~\eqref{eq:gap}, we can choose $2\gamma<\gamma_2'<\gamma_2$. For some $\beta_1>0$,
\begin{equation}
\begin{gathered}
e^{-iH_1t}=\sum_{a\in\Mslow}e^{-iz_at}x_ay_a^\dagger+\mathcal E(t),\\
\|\mathcal E\|+\|\dot{\mathcal E}\|\le Ke^{-(\gamma+\beta_1)t}.
\end{gathered}
\label{eq:app-slow-split}
\end{equation}
Also $\|e^{-iH_2t}\|+\|\partial_t e^{-iH_2t}\|\le Ke^{-\gamma_2't}$. Choose $0<\beta<\min\{\beta_1,\gamma_2'-2\gamma\}$. Let $Q(t)$ be the matrix in Eq.~\eqref{eq:general-Q} with its integral truncated at $t$. Positivity gives $0\preceq Q(t)\preceq Q$, and $Q-Q(t)$ decays exponentially through one derivative.

\paragraph{A differentiable factor on a fixed space.}
To control variation, define $\Lambda_t$ and $X=X_{ij}$ as in Eq.~\eqref{eq:balanced-pair-matrix}. As in Appendix~\ref{app:general-concurrence}, $\Ccal(X)=e^{2\gamma t}C_{ij}$.

Choose $L_{\rm in}$ with $\rho_{SM}(0)=L_{\rm in}L_{\rm in}^\dagger$. Write $\mathcal U(t)=\bigoplus_{n=0}^2e^{-iH_nt}$ and $\mathsf\Gamma_r=\sum_{n=1}^2\Gamma_r^{(n)}$. The amplitude for $k$ losses is
\begin{equation}
\Psi_k(t,u)=\mathcal U(t-u_k)\mathsf\Gamma_{r_k}\mathcal U(u_k-u_{k-1})
\cdots\mathsf\Gamma_{r_1}\mathcal U(u_1)L_{\rm in},
\label{eq:app-history}
\end{equation}
with $0<u_1<\cdots<u_k<t$ and $k\le2$. For $k=0$ it is $\mathcal U(t)L_{\rm in}$. Summing $\Psi_k\Psi_k^\dagger$ over labels and integrating over loss times gives the exact state. After the partial trace, retain the initial factor and traced configuration as column indices. Multiply the four pair rows by $e^{2\gamma t},e^{\gamma t},e^{\gamma t},1$. For each $k$, change variables to $u_l=ts_l$ and multiply the amplitudes by $t^{k/2}$. The resulting rows define a factor $L(t)$ on the fixed direct sum of the spaces $L^2(0<s_1<\cdots<s_k<1)$ with their finite column indices, where the $k=0$ summand is finite dimensional. The factor satisfies $X=LL^\dagger$.

This factor is continuously differentiable in Hilbert-Schmidt norm. Indeed, an amplitude $A(t,u)$ of the scaled rows becomes $G_t(s)=t^{k/2}A(t,ts)$, with
\begin{equation}
\dot G_t=t^{k/2}\left[\frac{k}{2t}A+\partial_tA
+\sum_{l=1}^k s_l\partial_{u_l}A\right]_{u=ts}.
\label{eq:app-fixed-derivative}
\end{equation}
For $\lambda t\ge1$, its squared Hilbert-Schmidt norm is bounded by a constant times the integral of $|A|^2+|\partial_tA|^2+\sum_l|\partial_{u_l}A|^2$ over the original loss times. Each partial derivative inserts a finite sector generator beside a propagator or jump, preserving the decay bounds. For one loss, the middle rows are bounded by $Ke^{-(\gamma_2'-\gamma)u}$ and the fourth row by $Ke^{-\gamma u}$. For two losses, only the fourth row survives, with bound $Ke^{-\gamma_2'u_1-\gamma(u_2-u_1)}$. These bounds are square integrable on the infinite ordered domains. The rows without loss are bounded, with first row $O(e^{-(\gamma_2'-2\gamma)t})$. Thus $L$ and $\dot L$ are bounded. 

Set the first row of $L$ to zero, retain in the middle rows only a final one-excitation state with its final propagator multiplied by $P_1$, keep the fourth row unchanged, and call the result $L_\star$. For a discarded middle row with one loss, its squared norm and that of its derivative obey
\begin{equation}
K e^{-2\beta_1t}\int_0^t
e^{-2(\gamma_2'-\gamma-\beta_1)u}\,du
\le Ke^{-2\beta t}.
\label{eq:app-factor-convolution}
\end{equation}
Equation~\eqref{eq:app-fixed-derivative} gives the derivative bound, and the strict choice of $\beta$ absorbs a possible factor of $t$. The other discarded terms decay at least as fast, so
\begin{equation}
\|L-L_\star\|_{\rm HS}+\|\dot L-\dot L_\star\|_{\rm HS}
\le Ke^{-\beta t}.
\label{eq:app-factor-c1}
\end{equation}
The retained middle rows have the form $\ell_k^\star=\sum_a v_a(k)e^{-i\omega_at}\zeta_a(t)$ for $k=i,j$, where $\zeta_a\zeta_b^\dagger=Q_{ab}(t)$. Different loss histories occupy orthogonal summands, so this Gram matrix adds the initial and feeding terms without interference between them. The inner product of the retained rows is therefore $b_\star=\Phi_{ij}+O(e^{-\beta t})$, also through one derivative.

\paragraph{Small singular values and the increment estimate.}
Put $W=L^\tp YL$ and $W_\star=L_\star^\tp YL_\star$. Equation~\eqref{eq:app-factor-c1} makes their difference exponentially small through one derivative. The two possibly nonzero singular values of $W_\star$ are
\begin{equation}
\xi_{1,2}^\star=\sqrt{a_\star c_\star}\pm|b_\star|,
\qquad a_\star=\|\ell_i^\star\|^2,
\quad c_\star=\|\ell_j^\star\|^2.
\label{eq:app-leading-roots}
\end{equation}
Work on late intervals where $|\Phi_{ij}|\ge m>0$. The first root and its gap from the second are then bounded below. For a nonbipartite graph or a pair in the same part, $v_a(i)/v_i=v_a(j)/v_j$ for each retained channel, so the two rows are parallel for every $Q(t)$ and the rank is one. For a pair in opposite parts, $a_\star c_\star-|b_\star|^2=4v_i^2v_j^2\det Q(t)$. If $\det Q=0$, the inequalities $0\preceq Q(t)\preceq Q$ again give rank one. If $\det Q>0$, the second root stays positive at late times. Thus the leading rank $r\in\{1,2\}$ is fixed and its nonzero roots have uniform spectral gaps.

Let $P$ and $P_\star$ project onto these $r$ right singular spaces. The resolvent identity for $W^\dagger W$ and $W_\star^\dagger W_\star$, differentiated once and integrated around their separated roots, gives $\xi_k-\xi_k^\star=O(e^{-\beta t})$ and $P-P_\star=O(e^{-\beta t})$, also through one derivative. The remaining singular values have sum $\|B_0\|_1$, where
\begin{equation}
\begin{split}
B_0&=W(\id-P)\\
&=(W-W_\star)(\id-P)+W_\star(P_\star-P).
\end{split}
\label{eq:app-small-root-operator}
\end{equation}
This identity bounds $B_0$ and $\dot B_0$ by $Ke^{-\beta t}$ in operator norm. Their ranks are at most four and eight, respectively, the latter because the derivative is a limit of differences of two rank-four operators. The same bounds therefore hold in trace norm.

The concurrence is positive on these intervals at late times. Using $\xi_1^\star-\xi_2^\star=2|b_\star|$ gives
\begin{equation}
\begin{gathered}
C_{ij}(t)=2e^{-2\gamma t}|\Phi_{ij}(t)|+f(t)-\|B(t)\|_1,\\
|f|+|\dot f|+\|B\|_1+\|\dot B\|_1
\le Ke^{-(2\gamma+\beta)t},
\end{gathered}
\label{eq:app-remainder-structure}
\end{equation}
where $B=e^{-2\gamma t}B_0$ and $f$ collects the smooth large root differences and $2e^{-2\gamma t}(|b_\star|-|\Phi_{ij}|)$. The inequality $|\|B(t+h)\|_1-\|B(t)\|_1|\le\|B(t+h)-B(t)\|_1$ now proves Eq.~\eqref{eq:remainder}.

\paragraph{Strict decrease and revivals.}
If the derivative of the leading concurrence is at most $-\delta e^{-2\gamma t}$, with $\delta>0$, Eq.~\eqref{eq:remainder} preserves this sign in every sufficiently late increment of the exact concurrence. This proves strict decrease for a nonzero constant $\Phi_{ij}$. For pairs in the same part, the maximal derivative coefficient after dividing by $2v_iv_j\Tr Qe^{-2\gamma t}$ is
\begin{equation}
\max_x\{-\Omega\nu\sin x-2\gamma(1+\nu\cos x)\}
=\nu\sqrt{\Omega^2+4\gamma^2}-2\gamma.
\label{eq:app-same-threshold}
\end{equation}
It is negative for $\nu<\nucrit$, and the profile is then bounded away from zero. For pairs in opposite parts with $U\ne0$, the maximal logarithmic derivative of the profile is $\Omega F^2/(2|U|\sqrt{U^2+F^2})$. Comparison with $2\gamma$ gives Eq.~\eqref{eq:cross-threshold} and strict decrease under the reverse inequality.

Above either threshold, each period contains separated intervals of decrease and increase for the leading concurrence. Hence both differences in Eq.~\eqref{eq:general-witness} are positive at $\tau_n=0$ for suitable $s_0<s_1<s_2$, and Appendix~\ref{app:asymptotic-laws} gives arbitrarily late revivals. Recurring zeros of the profile yield the same conclusion when $\nu=1$ or $U=0<F$.

With $x=\Omega t+\phi$, the coefficient maximized in Eq.~\eqref{eq:app-same-threshold} equals $-2\gamma-\nu\sqrt{\Omega^2+4\gamma^2}\cos(x-\vartheta)$. It changes sign from positive to negative at $x-\vartheta=-\chi$, which gives Eq.~\eqref{eq:times}. For pairs in opposite parts, put $x=\Omega t+\phi'$. The logarithmic derivative of the leading concurrence equals
\begin{equation*}
\frac{F^2\sqrt{\Omega^2+4\gamma^2}}{2(U^2+F^2\sin^2x)}\left[\cos(2x-\vartheta)-\nucrit\left(1+\frac{2U^2}{F^2}\right)\right],
\end{equation*}
which changes sign from positive to negative at $2x-\vartheta=\chi'$. This gives Eq.~\eqref{eq:times-cross}. The conditions $\nucrit<\nu$ and $F^2(1-\nucrit)>2\nucrit U^2$ make both zeros simple, and $|\Phi_{ij}|$ stays bounded away from zero near them. Write $t_n$ for a maximum in either equation. At $t_n$, the leading concurrence has a strictly negative second derivative proportional to $e^{-2\gamma t_n}$. In a fixed small neighborhood where this bound remains uniform, the derivative of the leading concurrence has magnitude at least $c e^{-2\gamma t_n}|t-t_n|$, with positive sign before $t_n$ and negative sign after it. Equation~\eqref{eq:remainder} then makes the exact concurrence strictly increasing before $t_n-Ke^{-\beta t_n}$ and strictly decreasing after $t_n+Ke^{-\beta t_n}$ within that neighborhood. By continuity, the intervening interval contains at least one local maximum, which proves the stated location bound.

\section{Numerical methods}
\label{app:numerics}

We construct the exact space with at most two total excitations, retaining bosonic square-root occupation factors and excluding qubit double occupation. We propagate the density matrix with the exponential of the sparse Lindblad generator using SciPy~\cite{Virtanen2020} and trace out all other qubits and memory modes to obtain the complete pair state. For the figures, the pair entries in the row and column of $|11\rangle$ come from the blocks $\rho^{(2,m)}$, which receive no loss feeding, and are evaluated directly from $e^{-iH_2t}$ and $e^{-iH_1t}$. Local extrema are located by bounded minimization and zeros of $\xi_1-\xi_2-\xi_3-\xi_4$ by root finding, with additional samples near each zero.

Near the vacuum, direct diagonalization of the product in Eq.~\eqref{eq:concurrence} can fail to resolve small populations accurately.
We instead set $d_a=\sqrt{\max(X_{aa},0)}$ and diagonalize the correlation matrix $K_{ab}=X_{ab}/(d_ad_b)$ on its nonzero diagonal support. If $K=U\operatorname{diag}(\kappa)U^\dagger$, restore $L=\operatorname{diag}(d)U\operatorname{diag}(\sqrt{\kappa})$. The Wootters roots are the singular values of $L^\tp YL$. Positivity sets a row and column to zero when its diagonal vanishes. 
Small negative eigenvalues of $K$, greater than $-10^{-8}$, are treated as numerical roundoff and set to zero, while more negative eigenvalues cause the calculation to terminate. We then verify the reconstructed matrix and check that it is positive semidefinite. All solid curves show the concurrence of the complete pair state.

The integral in Eq.~\eqref{eq:general-Q} is evaluated without fitting. For each retained pair of roots, solve
\begin{equation}
-i(H_2-z_a)\mathcal X_{ab}+i\mathcal X_{ab}(H_2^\dagger-\bar z_b)
=-\rho^{(2,2)}(0).
\label{eq:numerical-sylvester}
\end{equation}
Its contribution to $Q_{ab}$ is $\sum_r y_a^\dagger \Gamma_r^{(2)}\mathcal X_{ab}\Gamma_r^{(2)\dagger}y_b$. Before solving, we check the sufficient convergence condition $2\gamma_2>\gamma_a+\gamma_b$, where $\gamma_a=-\operatorname{Im}z_a$.

For Eq.~\eqref{eq:feeding-prep}, $Q=0.15Q^{(1)}+p_2Q^{(2)}$, where $Q^{(1)}$ has only $Q^{(1)}_{RR}=0.9933366153$. With $J_{11}=1$, unit population in $|13\rangle$ gives
\begin{equation}
\begin{aligned}
Q^{(2)}_{RR}=Q^{(2)}_{-R,-R}&=0.2011667011,\\
Q^{(2)}_{R,-R}&=0.0906937803-0.0000495797i.
\end{aligned}
\label{eq:feeding-matrix}
\end{equation}
Hermiticity fixes the other entry. These values give the stated crossings $p_2=0.29626396$ and $0.58152236$.

For Fig.~\ref{fig:times}, the state has no coherence between excitation numbers. The only nonzero off-diagonal entries of each pair matrix are then $X_{23}$ and $X_{32}$, so $C=2\max\{0,|X_{23}|-\sqrt{X_{11}X_{44}}\}$. Each exact maximum after the last zero of $C$ is a root of $dC/dt$, bracketed by a sign change on a grid of step $0.5/\lambda$. The evolution of the other entries of the full state does not involve its vacuum population. 
We therefore exclude the vacuum population during propagation and reconstruct $X_{44}$ from the unit trace, avoiding numerical errors dominated by the much larger vacuum contribution. A closed form built from the eigendecompositions of $H_1$ and $H_2$, evaluated in $40$-digit arithmetic, reproduces every computed maximum to within $10^{-11}/\lambda$.

An independent implementation using tensor-product operators reproduces the matrix elements obtained from the primary construction.
Density matrix integration with DOP853, at relative and absolute tolerances $2\times10^{-12}$ and $2\times10^{-14}$, agrees with sparse exponential propagation to within $1.5\times10^{-12}$ in the tested pair matrices, including those for the feeding preparation and the detuned $K_4$ state. A general $K_4$ test mixes $98\%$ of Eq.~\eqref{eq:critical-preparation} at $p=0.01$ with $2\%$ of a random density operator on $\mathcal Q_0\oplus\mathcal Q_1\oplus\mathcal Q_2$, generated from an $11\times11$ complex Gaussian matrix with NumPy seed $20260923$. Every cross-sector coherence block is nonzero. The calculation retains all least damped modes and includes the $B_{21}$ contribution to $K_{10}$. Quadrature and Sylvester evaluations agree within $1.4\times10^{-12}$. At $\lambda t=100$, $X_{12}(t)$ differs from $X_{12}^{(0)}(t)$ by less than $2.2\times10^{-9}$ entrywise.

Additional checks cover $200$ random pair matrices, the exact equal-rate preparation at $p=0.01,1/2$, repeated channel roots, the example with $Q_{RR}=0$, and both strict sides of each bipartite threshold. The latter two tests use $g/\lambda=0.3,1,2$.

The spectral checks cover $56$ resonant cases on seven graphs, the exact detuned $K_4$ equality point, and $276$ additional parameter choices spanning resonance and nonzero detuning, up to $g/\lambda=30$. The smallest computed value of $\gamma_2-2\gamma_1$ at nonzero detuning is above $-10^{-14}\lambda$, within roundoff, and selected $H_2$ eigenpair residuals are below $10^{-14}$ relative to the matrix norm.

\bibliography{references}

\end{document}

%% file: sector_rate_bound.tex
\section{Sector decay bound and its equality case}
\label{sector-bound-proof}

We prove Proposition~\ref{prop:sector-rate-bound} for $g,\eta,\lambda>0$ and $\Delta\in\mathbb R$. For a Hermitian matrix $X$, we write $X\succeq0$ if $X$ is positive semidefinite and $X\succ0$ if $X$ is positive definite. Section~\ref{sec:model} and Appendix~\ref{app:slow-detuning} show that $0<\gamma_1<\lambda/2$ and $|D_\theta|=\max_{\vartheta\in\spec A}|D_\vartheta|$ for every $\theta\in\mathcal M$. Define
\begin{equation}
\tau=\lambda-2\gamma_1,\qquad
d=\frac{\max_{\theta\in\spec A}D_\theta^2}{\tau^2}.
\label{sector-bound-normalization}
\end{equation}
Then $\tau,d>0$, and for $\theta\in\mathcal M$, substituting $z_\theta$ into its channel equation gives
\begin{equation}
g^2=\gamma_1(\lambda-\gamma_1)(1+d).
\label{sector-bound-coupling}
\end{equation}

\subsection{Proof of the sector decay bound}

Introduce auxiliary bosonic modes $b_r$ by replacing $\sigma_r^-$ and $\sigma_r^+$ with $b_r$ and $b_r^\dagger$, where $[b_r,b_s^\dagger]=\delta_{rs}$, and leave the memory modes unchanged. The auxiliary one-excitation generator is still $H_1$. Since the auxiliary model has no interactions between excitations, its two-excitation generator is
\begin{equation*}
\widetilde H_2
=\left.(H_1\otimes\id+\id\otimes H_1)\right|_{\operatorname{Sym}^2\mathcal H_1},
\qquad \mathcal H_1=\mathbb C^{2N},
\end{equation*}
where $\operatorname{Sym}^2\mathcal H_1$ is the subspace spanned by $u\otimes v+v\otimes u$ with $u,v\in\mathcal H_1$. Let $\mathsf E:\operatorname{ran}\Pi_2\to\operatorname{Sym}^2\mathcal H_1$ be the linear map that replaces $\sigma_r^+$ with $b_r^\dagger$. Then $\mathsf E^\dagger\mathsf E=\id$, and $\operatorname{ran}\mathsf E$ is the orthogonal complement of the $N$ states $(b_r^\dagger)^2|0\rangle/\sqrt2$. Let $\mathsf U=\mathsf E|_{\mathcal Q_2\otimes|\Omega_M\rangle}$. For simplicity, we denote $\mathsf E\psi$ by $\psi$. In the decomposition by zero or positive memory occupation,
\begin{equation*}
\widetilde H_2=
\begin{pmatrix}
0&\mathsf V^\dagger\\
\mathsf V&\mathsf K
\end{pmatrix},
\qquad
H_2=
\begin{pmatrix}
0&\mathsf U^\dagger\mathsf V^\dagger\\
\mathsf V\mathsf U&\mathsf K
\end{pmatrix}.
\end{equation*}
Thus $H_2=\mathsf E^\dagger\widetilde H_2\mathsf E$.

Let $e_r$ and $f_r$ be the vectors $\sigma_r^+|0\rangle$ and $m_r^\dagger|0\rangle$ of $\mathcal B_1$, and put $e_{rr}=e_r\otimes e_r$. Then $e_{rr}=(b_r^\dagger)^2|0\rangle/\sqrt2$, so $e_{rr}^\dagger\psi=0$ for $\psi\in\operatorname{ran}\mathsf E$. We call $e_{rr}^\dagger\psi$ the qubit double-occupation amplitude of $\psi\in\operatorname{Sym}^2\mathcal H_1$ at vertex $r$. By the zero block of $\widetilde H_2$, we have $e_{rr}^\dagger\widetilde H_2e_{ss}=0$. Since $H_1e_r=H_1^\dagger e_r=gf_r$,
\begin{equation}
\widetilde H_2e_{rr}=\widetilde H_2^\dagger e_{rr}=g(e_r\otimes f_r+f_r\otimes e_r).
\label{sector-bound-leak}
\end{equation}

Define
\begin{equation}
\ell=\lambda(1+d),\qquad
\kappa=\frac{2g}{\ell},\qquad
\mathsf W=\begin{pmatrix}1&-i\kappa\\ i\kappa&1\end{pmatrix}.
\label{sector-bound-metric}
\end{equation}
Equation~\eqref{sector-bound-coupling} gives $g\kappa=2\gamma_1(\lambda-\gamma_1)/\lambda$ and $\kappa^2=4\gamma_1(\lambda-\gamma_1)/[\lambda^2(1+d)]$. Since $4\gamma_1(\lambda-\gamma_1)\le\lambda^2$ and $d>0$, we have $\kappa<1$, which implies $\mathsf W\succ0$. For real $D$, let $h_D$ be the matrix in Eq.~\eqref{eq:channel-generator} with $D_\theta$ replaced by $D$. Direct multiplication gives
\begin{equation}
\begin{aligned}
&-i\bigl(h_D^\dagger\mathsf W-\mathsf Wh_D\bigr)-2\gamma_1\mathsf W=\mathsf Q_D,\\
&\mathsf Q_D=\begin{pmatrix}
2\gamma_1\tau/\lambda&\kappa(D-i\tau)\\
\kappa(D+i\tau)&2(\lambda-\gamma_1)\tau/\lambda
\end{pmatrix}.
\end{aligned}
\label{sector-bound-channel}
\end{equation}
The diagonal entries of $\mathsf Q_D$ are positive. By the expression for $\kappa^2$, we have $\det \mathsf Q_D=\kappa^2(d\tau^2-D^2)$. Since $D_\theta^2\le d\tau^2$, every $\mathsf Q_{D_\theta}\succeq0$.

In the basis $\mathcal B_1$, Eqs.~\eqref{eq:H1} and~\eqref{eq:adjacency-decomposition} give $H_1=\sum_\theta h_{D_\theta}\otimes E_\theta$, where the first factor acts on the qubit and memory blocks. With $\mathsf W_1=\mathsf W\otimes\id_N$ and $\mathsf Q_1=\sum_\theta \mathsf Q_{D_\theta}\otimes E_\theta\succeq0$,
\begin{equation*}
-i\bigl(H_1^\dagger \mathsf W_1-\mathsf W_1H_1\bigr)-2\gamma_1\mathsf W_1=\mathsf Q_1.
\end{equation*}
Restricted to $\operatorname{Sym}^2\mathcal H_1$, the operators $\mathsf W_2=\mathsf W_1\otimes \mathsf W_1\succ0$ and $\mathsf Q_2=\mathsf Q_1\otimes \mathsf W_1+\mathsf W_1\otimes \mathsf Q_1\succeq0$ satisfy
\begin{equation}
-i\bigl(\widetilde H_2^\dagger \mathsf W_2-\mathsf W_2\widetilde H_2\bigr)-4\gamma_1\mathsf W_2=\mathsf Q_2.
\label{sector-bound-two}
\end{equation}
Using $\mathsf W_1e_r=e_r+i\kappa f_r$, $\mathsf W_1f_r=f_r-i\kappa e_r$, $\kappa\ell=2g$, and Eq.~\eqref{sector-bound-leak}, both sides of
\begin{equation}
\mathsf W_2(\widetilde H_2+i\ell)e_{rr}=-(1-\kappa^2)(\widetilde H_2^\dagger-i\ell)e_{rr}
\label{sector-bound-compatible}
\end{equation}
equal $(1-\kappa^2)[i\ell e_{rr}-g(e_r\otimes f_r+f_r\otimes e_r)]$.

Let $H_2\psi=z\psi$ with $\psi\ne0$. Then $e_{rr}^\dagger\psi=0$, and the block forms give
\begin{equation*}
\widetilde H_2\psi=z\psi+\sum_rc_re_{rr},\qquad c_r=e_{rr}^\dagger\widetilde H_2\psi.
\end{equation*}
Put $\phi=(\widetilde H_2+i\ell)\psi$. Using $e_{rr}^\dagger\psi=0$ and $e_{rr}^\dagger\widetilde H_2e_{ss}=0$,
\begin{equation*}
\begin{aligned}
(z-\widetilde H_2)\phi&=-\sum_rc_r(\widetilde H_2+i\ell)e_{rr},\\
e_{rr}^\dagger(\widetilde H_2+i\ell)\phi&=(z+2i\ell)c_r.
\end{aligned}
\end{equation*}
Equation~\eqref{sector-bound-compatible} then yields
\begin{equation*}
\phi^\dagger \mathsf W_2(z-\widetilde H_2)\phi=(1-\kappa^2)(\bar z-2i\ell)\sum_r|c_r|^2.
\end{equation*}
By Eq.~\eqref{sector-bound-two}, the imaginary part of the left side is $(\operatorname{Im}z+2\gamma_1)\phi^\dagger \mathsf W_2\phi+\tfrac12\phi^\dagger \mathsf Q_2\phi$. Therefore
\begin{equation}
\begin{aligned}
0={}&(\operatorname{Im}z+2\gamma_1)\phi^\dagger \mathsf W_2\phi+\tfrac12\phi^\dagger \mathsf Q_2\phi\\
&+(1-\kappa^2)(\operatorname{Im}z+2\ell)\sum_r|c_r|^2.
\end{aligned}
\label{sector-bound-energy}
\end{equation}
Suppose $\operatorname{Im}z\ge-2\gamma_1$. Since $\ell>\lambda>2\gamma_1$ and $\kappa<1$, all three terms are nonnegative, so each vanishes. Thus every $c_r$ vanishes and $\widetilde H_2\psi=z\psi$. If $\operatorname{Im}z>-2\gamma_1$, the first term also gives $\phi=0$. Since $\phi=(z+i\ell)\psi$ and $\operatorname{Im}(z+i\ell)>0$, this implies $\psi=0$, a contradiction. No eigenvalue of $H_2$ has imaginary part greater than $-2\gamma_1$, so $\gamma_2\ge2\gamma_1$.

\subsection{The equality case}

\begin{lemma}[Equality criterion]\label{sector-bound-equality}
Equality $\gamma_2=2\gamma_1$ holds if and only if there exist $\psi\ne0$ and $z$ such that $\widetilde H_2\psi=z\psi$, $\operatorname{Im}z=-2\gamma_1$, and $e_{rr}^\dagger\psi=0$ for every $r\in V$.
\end{lemma}

For necessity, let $H_2\psi=z\psi$ with $\psi\ne0$ and $\operatorname{Im}z=-2\gamma_1$. Equation~\eqref{sector-bound-energy} gives $c_r=0$ for every $r$, so $\psi$ is an eigenvector of $\widetilde H_2$ with zero qubit double-occupation amplitudes. Conversely, any auxiliary eigenvector satisfying these conditions lies in $\operatorname{ran}\mathsf E$ and is an eigenvector of $H_2$. The sector decay bound then gives equality.

\begin{lemma}\label{sector-bound-semisimple}
If $z\in\spec H_2$ and $\operatorname{Im}z=-2\gamma_1$, then $z$ is semisimple, that is, every Jordan block of $H_2$ at $z$ has size one.
\end{lemma}

Let $\operatorname{Im}z=-2\gamma_1+\varepsilon$ and $\xi\in\operatorname{ran}\Pi_2$, where $0<\varepsilon\le\lambda$. By the sector decay bound, $\psi=(z-H_2)^{-1}\xi$ exists. Put $c_r=e_{rr}^\dagger\widetilde H_2\psi$ and $\phi=(\widetilde H_2+i\ell)\psi$. The block forms give $\widetilde H_2\psi=z\psi-\xi+\sum_rc_re_{rr}$. Using $e_{rr}^\dagger\xi=e_{rr}^\dagger\psi=0$ and $e_{rr}^\dagger\widetilde H_2e_{ss}=0$,
\begin{equation*}
\begin{aligned}
(z-\widetilde H_2)\phi&=(\widetilde H_2+i\ell)\Bigl(\xi-\sum_rc_re_{rr}\Bigr),\\
e_{rr}^\dagger(\widetilde H_2+i\ell)\phi&=(z+2i\ell)c_r-e_{rr}^\dagger\widetilde H_2\xi.
\end{aligned}
\end{equation*}
As in the derivation of Eq.~\eqref{sector-bound-energy}, Eqs.~\eqref{sector-bound-two} and~\eqref{sector-bound-compatible} give
\begin{equation}
\begin{aligned}
&\varepsilon\,\phi^\dagger \mathsf W_2\phi+\tfrac12\phi^\dagger \mathsf Q_2\phi+(1-\kappa^2)(\operatorname{Im}z+2\ell)\|c\|^2\\
&=\operatorname{Im}\bigl[\phi^\dagger \mathsf W_2(\widetilde H_2+i\ell)\xi\bigr]
-(1-\kappa^2)\operatorname{Im}\sum_rc_r\overline{e_{rr}^\dagger\widetilde H_2\xi},
\end{aligned}
\label{sector-bound-resolvent}
\end{equation}
where $\|c\|^2=\sum_r|c_r|^2$. The right side is at most a constant times $\|\xi\|(\|\phi\|+\|c\|)$. Since $\mathsf W_2\succ0$, $\mathsf Q_2\succeq0$, $\kappa<1$, $\operatorname{Im}z+2\ell\ge2(\ell-\gamma_1)>0$, and $\varepsilon\le\lambda$, the left side is at least a positive constant times $\varepsilon(\|\phi\|+\|c\|)^2$. Both constants are independent of $z$ and $\xi$, so $\|\phi\|+\|c\|=O(\|\xi\|/\varepsilon)$. Since $\phi=(z+i\ell)\psi-\xi+\sum_rc_re_{rr}$ and $|z+i\ell|\ge\ell-2\gamma_1>0$, this gives $\|(z-H_2)^{-1}\|=O(\varepsilon^{-1})$ as $\varepsilon\to0$. If $H_2$ had a Jordan block of size $m\ge2$ at $z_0$ with $\operatorname{Im}z_0=-2\gamma_1$, then $\|(z_0+i\varepsilon-H_2)^{-1}\|$ would grow like $\varepsilon^{-m}$, a contradiction.

For $\theta\in\mathcal M$, write $z_\theta=\omega_\theta-i\gamma_1$. Taking the imaginary part of the channel equation, we have $\omega_\theta=-\gamma_1D_\theta/\tau$. Since $D_\theta\ne0$, we have $\omega_\theta\ne0$. At resonance,
\begin{equation}
\begin{aligned}
z_R&=\omega_R-i\gamma_1,&
\omega_R&=-\frac{\gamma_1\eta R}{\lambda-2\gamma_1}<0,\\
z_{-R}&=-\omega_R-i\gamma_1,
\end{aligned}
\label{sector-bound-resonance}
\end{equation}
where the second root is present only for a bipartite graph. For $\theta\in\mathcal M$, Appendix~\ref{app:slow-detuning} shows that $z_\theta$ is a semisimple eigenvalue of $H_1$ with eigenvectors
\begin{equation}
\chi_u=\begin{pmatrix}u\\z_\theta u/g\end{pmatrix},\qquad Au=\theta u,\quad u\ne0.
\label{sector-bound-slow-vectors}
\end{equation}

By triangularizing $H_1$, we have $\spec\widetilde H_2\subseteq\{z+z':z,z'\in\spec H_1\}$. If $-\operatorname{Im}(z+z')=2\gamma_1$, then $-\operatorname{Im}z=-\operatorname{Im}z'=\gamma_1$. Since these summands are semisimple, every eigenvector of $\widetilde H_2$ with decay rate $2\gamma_1$ is a sum of symmetric products $\chi_u\otimes\chi_{u'}+\chi_{u'}\otimes\chi_u$ for which the eigenvalues of $\chi_u$ and $\chi_{u'}$ add to its eigenvalue. The qubit double-occupation amplitude of such a product at vertex $r$ is $2u_ru'_r$.

If $\mathcal M=\{\theta\}$, the only auxiliary eigenvalue with decay rate $2\gamma_1$ is $2z_\theta$. Let $\{u_k\}$ be a basis of the eigenspace of $A$ at $\theta$. The eigenvectors at $2z_\theta$ are the vectors $\psi=\sum_{k,l}a_{kl}\chi_{u_k}\otimes\chi_{u_l}$ with $a_{kl}=a_{lk}$ not all zero, and the qubit double-occupation amplitudes of $\psi$ are the diagonal entries of
\begin{equation}
S=\sum_{k,l}a_{kl}u_ku_l^\tp.
\label{sector-bound-S}
\end{equation}
The map from the coefficients $a_{kl}$ to $S$ is a bijection onto the symmetric matrices with $AS=\theta S$. For $\theta=R$, we have $S=cvv^\tp$ with $c\ne0$, so $S_{rr}=cv_r^2\ne0$.

If $\mathcal M=\{R,\theta_-\}$, then $z_{\theta_-}=-\overline{z_R}$, and the auxiliary eigenvalues with decay rate $2\gamma_1$ are $2z_R$, $z_R+z_{\theta_-}$, and $2z_{\theta_-}$. Their real parts $2\omega_R$, $0$, and $-2\omega_R$ are distinct, so each eigenvector belongs to one of the three eigenspaces. The eigenspaces at $2z_R$ and $2z_{\theta_-}$ are described by Eq.~\eqref{sector-bound-S} with $\theta=R$ and $\theta=\theta_-$. The eigenspace at $z_R+z_{\theta_-}$ consists of the vectors $\chi_v\otimes\chi_u+\chi_u\otimes\chi_v$ with $Au=\theta_-u$. Their qubit double-occupation amplitudes $2v_ru_r$ vanish at every vertex only if $u=0$.

By Eq.~\eqref{eq:general-slow-labels}, $\theta_-\in\mathcal M$ if and only if $\Delta\le-\eta(R+\theta_-)/2$. Combining this with Lemma~\ref{sector-bound-equality}, we have $\gamma_2=2\gamma_1$ if and only if, in addition, $AS=\theta_-S$ for some nonzero symmetric $S$ with $S_{rr}=0$. If $\theta_-$ is simple with eigenvector $u$, then $S=cuu^\tp$ with $c\ne0$, and the entries $S_{rr}=cu_r^2$ do not all vanish. Since $R+\theta_-\ge0$, if $\Delta\ge0$ then either $\Delta>-\eta(R+\theta_-)/2$, or $R+\theta_-=0$ and the graph is bipartite with $\theta_-=-R$ simple. This proves Proposition~\ref{prop:sector-rate-bound}.

\begin{proposition}\label{sector-bound-bipartite-detuning}
For a connected bipartite graph, $\gamma_2>2\gamma_1$ for every $\Delta\in\mathbb R$ and every finite $g,\eta,\lambda>0$.
\end{proposition}

This follows from Proposition~\ref{prop:sector-rate-bound}, because $\theta_-=-R$ is simple.

%% file: references.bib
@article{Bellomo2007,
  author = {Bruno Bellomo and Lo Franco, Rosario and Giuseppe Compagno},
  title = {{Non-Markovian effects on the dynamics of entanglement}},
  journal = {Phys. Rev. Lett.},
  volume = {99},
  pages = {160502},
  year = {2007},
  doi = {10.1103/PhysRevLett.99.160502},
}

@article{YuEberly2004,
  author = {Ting Yu and J. H. Eberly},
  title = {Finite-time disentanglement via spontaneous emission},
  journal = {Phys. Rev. Lett.},
  volume = {93},
  pages = {140404},
  year = {2004},
  doi = {10.1103/PhysRevLett.93.140404},
}

@article{Aolita2015,
 author={Aolita, Leandro and {de Melo}, Fernando and Davidovich, Luiz},
 title={{Open-system dynamics of entanglement: a key issues review}},
 journal={Rep. Prog. Phys.}, volume={78}, pages={042001}, year={2015},
 doi={10.1088/0034-4885/78/4/042001}
}

@article{Garraway1997,
 author={Garraway, B. M.},
 title={{Nonperturbative decay of an atomic system in a cavity}},
 journal={Phys. Rev. A}, volume={55}, pages={2290--2303}, year={1997},
 doi={10.1103/PhysRevA.55.2290}
}

@article{Tamascelli2018,
 author={Tamascelli, D. and Smirne, A. and Huelga, S. F. and Plenio, M. B.},
 title={{Nonperturbative treatment of non-Markovian dynamics of open quantum systems}},
 journal={Phys. Rev. Lett.}, volume={120}, pages={030402}, year={2018},
 doi={10.1103/PhysRevLett.120.030402}
}

@article{Wootters1998,
  author = {William K. Wootters},
  title = {Entanglement of formation of an arbitrary state of two qubits},
  journal = {Phys. Rev. Lett.},
  volume = {80},
  pages = {2245--2248},
  year = {1998},
  doi = {10.1103/PhysRevLett.80.2245},
}

@article{Nokkala2016,
 author={Nokkala, Johannes and Galve, Fernando and Zambrini, Roberta and Maniscalco, Sabrina and Piilo, Jyrki},
 title={{Complex quantum networks as structured environments: engineering and probing}},
 journal={Sci. Rep.}, volume={6}, pages={26861}, year={2016},
 doi={10.1038/srep26861}
}

@article{Pleasance2020,
 author={Pleasance, Graeme and Garraway, Barry M. and Petruccione, Francesco},
 title={{Generalized theory of pseudomodes for exact descriptions of non-Markovian quantum processes}},
 journal={Phys. Rev. Res.}, volume={2}, pages={043058}, year={2020},
 doi={10.1103/PhysRevResearch.2.043058}
}

@article{Medina2021,
 author={Medina, Ivan and Garc{\'\i}a-Vidal, Francisco J. and Fern{\'a}ndez-Dom{\'\i}nguez, Antonio I. and Feist, Johannes},
 title={{Few-mode field quantization of arbitrary electromagnetic spectral densities}},
 journal={Phys. Rev. Lett.}, volume={126}, pages={093601}, year={2021},
 doi={10.1103/PhysRevLett.126.093601}
}

@article{Huang2026,
 author={Huang, Zhen and Park, Gunhee and Chan, Garnet Kin-Lic and Lin, Lin},
 title={{Coupled Lindblad pseudomode theory for simulating open quantum systems}},
 journal={Phys. Rev. Lett.}, volume={136}, pages={090403}, year={2026},
 doi={10.1103/qfkp-jc7s}
}

@article{SchmidtKoch2013,
  author = {Sebastian Schmidt and Jens Koch},
  title = {{Circuit QED lattices: Towards quantum simulation with superconducting circuits}},
  journal = {Ann. Phys. (Berlin)},
  volume = {525},
  pages = {395--412},
  year = {2013},
  doi = {10.1002/andp.201200261},
}

@article{Kollar2019,
  author = {Alicia J. Koll{\'a}r and Mattias Fitzpatrick and Andrew A. Houck},
  title = {{Hyperbolic lattices in circuit quantum electrodynamics}},
  journal = {Nature},
  volume = {571},
  pages = {45--50},
  year = {2019},
  doi = {10.1038/s41586-019-1348-3},
}

@article{Sundaresan2019,
 author={Sundaresan, Neereja M. and Lundgren, Rex and Zhu, Guanyu and Gorshkov, Alexey V. and Houck, Andrew A.},
 title={{Interacting qubit-photon bound states with superconducting circuits}},
 journal={Phys. Rev. X}, volume={9}, pages={011021}, year={2019},
 doi={10.1103/PhysRevX.9.011021}
}

@article{Ferreira2021,
 author={Ferreira, Vinicius S. and Banker, Jash and Sipahigil, Alp and Matheny, Matthew H. and Keller, Andrew J. and Kim, Eunjong and Mirhosseini, Mohammad and Painter, Oskar},
 title={{Collapse and revival of an artificial atom coupled to a structured photonic reservoir}},
 journal={Phys. Rev. X}, volume={11}, pages={041043}, year={2021},
 doi={10.1103/PhysRevX.11.041043}
}

@article{Man2015,
  author = {Zhong-Xiao Man and Yun-Jie Xia and Lo Franco, Rosario},
  title = {{Cavity-based architecture to preserve quantum coherence and entanglement}},
  journal = {Sci. Rep.},
  volume = {5},
  pages = {13843},
  year = {2015},
  doi = {10.1038/srep13843},
}

@article{Basit2018,
  author = {Abdul Basit and Hamad Ali and Fazal Badshah and Haiyang Zhang and Rafi Ud Din and Guo-Qin Ge},
  title = {{Protection of entanglement between two qubits trapped in coupled cavities via cavity architecture}},
  journal = {Laser Phys. Lett.},
  volume = {15},
  pages = {115204},
  year = {2018},
  doi = {10.1088/1612-202X/aae137},
}

@article{Roccati2022,
  author = {Federico Roccati and Salvatore Lorenzo and Giuseppe Calaj{\`o} and G. Massimo Palma and Angelo Carollo and Francesco Ciccarello},
  title = {{Exotic interactions mediated by a non-Hermitian photonic bath}},
  journal = {Optica},
  volume = {9},
  pages = {565},
  year = {2022},
  doi = {10.1364/OPTICA.443955},
}

@article{Roccati2024,
  author = {Federico Roccati and Miguel Bello and Zongping Gong and Masahito Ueda and Francesco Ciccarello and Aur{\'e}lia Chenu and Angelo Carollo},
  title = {{Hermitian and non-Hermitian topology from photon-mediated interactions}},
  journal = {Nat. Commun.},
  volume = {15},
  pages = {2400},
  year = {2024},
  doi = {10.1038/s41467-024-46471-w},
}

@article{Makin2009,
  author = {M. I. Makin and Jared H. Cole and Charles D. Hill and Andrew D. Greentree and Lloyd C. L. Hollenberg},
  title = {{Time evolution of the one-dimensional Jaynes-Cummings-Hubbard Hamiltonian}},
  journal = {Phys. Rev. A},
  volume = {80},
  pages = {043842},
  year = {2009},
  doi = {10.1103/PhysRevA.80.043842},
}

@article{Mendonca2020,
  author = {J. P. Mendon{\c{c}}a and F. A. B. F. de Moura and M. L. Lyra and G. M. A. Almeida},
  title = {{Generation and distribution of atomic entanglement in coupled-cavity arrays}},
  journal = {Phys. Rev. A},
  volume = {102},
  pages = {062416},
  year = {2020},
  doi = {10.1103/PhysRevA.102.062416},
}

@article{Zurita2025,
  author = {Zurita, Juan and Agust{\'i} Casado, Andr{\'e}s and Creffield, Charles E. and Platero, Gloria},
  title = {Multipartite entanglement distribution in a topological photonic network},
  journal = {Quantum},
  volume = {9},
  pages = {1625},
  year = {2025},
  doi = {10.22331/q-2025-02-10-1625}
}

@article{DiFidio2024,
  author = {Di Fidio, Christian and Ares, Laura and Sperling, Jan},
  title = {{Quantum walks and entanglement in cavity networks}},
  journal = {Phys. Rev. A},
  volume = {110},
  pages = {013705},
  year = {2024},
  doi = {10.1103/PhysRevA.110.013705},
}

@article{Li2015,
 author={C. Li and X. Z. Zhang and Z. Song},
 title={Equivalent spin-orbit interaction in the two-polariton {Jaynes-Cummings-Hubbard} model},
 journal={Sci. Rep.}, volume={5}, pages={11945}, year={2015},
 doi={10.1038/srep11945}
}

@article{WongLaw2011,
 author={Wong, Max T. C. and Law, C. K.},
 title={{Two-polariton bound states in the Jaynes-Cummings-Hubbard model}},
 journal={Phys. Rev. A}, volume={83}, pages={055802}, year={2011},
 doi={10.1103/PhysRevA.83.055802}
}

@article{Memarzadeh2013,
  author = {Laleh Memarzadeh and Stefano Mancini},
  title = {{Entanglement dynamics for qubits dissipating into a common environment}},
  journal = {Phys. Rev. A},
  volume = {87},
  pages = {032303},
  year = {2013},
  doi = {10.1103/PhysRevA.87.032303},
}

@article{Albrecht2019,
 author={Albrecht, Andreas and Henriet, Lo{\"\i}c and Asenjo-Garcia, Ana and Dieterle, Paul B. and Painter, Oskar and Chang, Darrick E.},
 title={{Subradiant states of quantum bits coupled to a one-dimensional waveguide}},
 journal={New J. Phys.}, volume={21}, pages={025003}, year={2019},
 doi={10.1088/1367-2630/ab0134}
}

@article{ZhangMolmer2019,
 author={Yu-Xiang Zhang and Klaus M{\o}lmer},
 title={Theory of Subradiant States of a One-Dimensional Two-Level Atom Chain},
 journal={Phys. Rev. Lett.}, volume={122}, pages={203605}, year={2019},
 doi={10.1103/PhysRevLett.122.203605}
}

@article{Poshakinskiy2021,
 author={Poshakinskiy, Alexander V. and Poddubny, Alexander N.},
 title={{Dimerization of many-body subradiant states in waveguide quantum electrodynamics}},
 journal={Phys. Rev. Lett.}, volume={127}, pages={173601}, year={2021},
 doi={10.1103/PhysRevLett.127.173601}
}

@article{ZhangYuMolmer2020,
 author={Zhang, Yu-Xiang and Yu, Chuan and M{\o}lmer, Klaus},
 title={{Subradiant bound dimer excited states of emitter chains coupled to a one dimensional waveguide}},
 journal={Phys. Rev. Res.}, volume={2}, pages={013173}, year={2020},
 doi={10.1103/PhysRevResearch.2.013173}
}

@article{TanasFicek2004,
  author = {R. Tana{\'s} and Z. Ficek},
  title = {{Entangling two atoms via spontaneous emission}},
  journal = {J. Opt. B: Quantum Semiclass. Opt.},
  volume = {6},
  pages = {S90--S97},
  year = {2004},
  doi = {10.1088/1464-4266/6/3/015},
}

@article{Mazzola2009,
 author={Mazzola, L. and Maniscalco, S. and Piilo, J. and Suominen, K.-A. and Garraway, B. M.},
 title={{Sudden death and sudden birth of entanglement in common structured reservoirs}},
 journal={Phys. Rev. A}, volume={79}, pages={042302}, year={2009},
 doi={10.1103/PhysRevA.79.042302}
}

@article{Ficek2006,
  author = {Z. Ficek and R. Tana{\'s}},
  title = {{Dark periods and revivals of entanglement in a two-qubit system}},
  journal = {Phys. Rev. A},
  volume = {74},
  pages = {024304},
  year = {2006},
  doi = {10.1103/PhysRevA.74.024304},
}

@article{Das2009,
  author = {Sumanta Das and G. S. Agarwal},
  title = {{Bright and dark periods in the entanglement dynamics of interacting qubits in contact with the environment}},
  journal = {J. Phys. B: At. Mol. Opt. Phys.},
  volume = {42},
  pages = {141003},
  year = {2009},
  doi = {10.1088/0953-4075/42/14/141003},
}

@article{Greentree2006,
  author = {Andrew D. Greentree and Charles Tahan and Jared H. Cole and Lloyd C. L. Hollenberg},
  title = {{Quantum phase transitions of light}},
  journal = {Nat. Phys.},
  volume = {2},
  pages = {856--861},
  year = {2006},
  doi = {10.1038/nphys466},
}

@article{Angelakis2007,
  author = {Dimitris G. Angelakis and Marcelo F. Santos and Sougato Bose},
  title = {{Photon-blockade-induced Mott transitions and XY spin models in coupled cavity arrays}},
  journal = {Phys. Rev. A},
  volume = {76},
  pages = {031805(R)},
  year = {2007},
  doi = {10.1103/PhysRevA.76.031805},
}

@article{Lindblad1976,
 author={G. Lindblad},
 title={On the generators of quantum dynamical semigroups},
 journal={Commun. Math. Phys.}, volume={48}, pages={119--130}, year={1976},
 doi={10.1007/BF01608499}
}

@article{GoriniKossakowskiSudarshan1976,
 author={Vittorio Gorini and Andrzej Kossakowski and E. C. G. Sudarshan},
 title={Completely positive dynamical semigroups of {N}-level systems},
 journal={J. Math. Phys.}, volume={17}, pages={821--825}, year={1976},
 doi={10.1063/1.522979}
}

@article{Torres2014,
 author={Juan Mauricio Torres},
 title={Closed-form solution of {Lindblad} master equations without gain},
 journal={Phys. Rev. A}, volume={89}, pages={052133}, year={2014},
 doi={10.1103/PhysRevA.89.052133}
}

@book{BrouwerHaemers2012,
 author={Andries E. Brouwer and Willem H. Haemers},
 title={Spectra of Graphs},
 publisher={Springer},
 address={New York},
 year={2012},
 doi={10.1007/978-1-4614-1939-6}
}

@misc{HashemiRafsanjani2012,
  author = {Hashemi Rafsanjani, S. M. and Agarwal, S.},
  title = {{X}-matrices provide a lower bound of concurrence},
  year = {2012},
  eprint = {1204.3912},
  archivePrefix = {arXiv},
  primaryClass = {quant-ph},
  doi = {10.48550/arXiv.1204.3912}
}

@book{HornJohnson2013,
  author = {Horn, Roger A. and Johnson, Charles R.},
  title = {Matrix Analysis},
  edition = {2},
  publisher = {Cambridge University Press},
  address = {Cambridge},
  year = {2013},
}

@article{Verstraete2001,
  author = {Frank Verstraete and Jeroen Dehaene and {De Moor}, Bart},
  title = {{Local filtering operations on two qubits}},
  journal = {Phys. Rev. A},
  volume = {64},
  pages = {010101(R)},
  year = {2001},
  doi = {10.1103/PhysRevA.64.010101},
}

@article{PowersStormer1970,
 author={Robert T. Powers and Erling St{\o}rmer},
 title={Free states of the canonical anticommutation relations},
 journal={Commun. Math. Phys.}, volume={16}, pages={1--33}, year={1970},
 doi={10.1007/BF01645492}
}

@article{Virtanen2020,
  author = {Pauli Virtanen and Ralf Gommers and Travis E. Oliphant and others},
  title = {{SciPy} 1.0: fundamental algorithms for scientific computing in {Python}},
  journal = {Nat. Methods},
  volume = {17},
  pages = {261--272},
  year = {2020},
  doi = {10.1038/s41592-019-0686-2},
}

@article{Mimar2019,
  author  = {Mimar, Sayat and Mussa Juane, Mariamo and Park, Juyong and Mu{\~n}uzuri, Alberto P. and Ghoshal, Gourab},
  title   = {Turing patterns mediated by network topology in homogeneous active systems},
  journal = {Phys. Rev. E},
  volume  = {99},
  pages   = {062303},
  year    = {2019},
  doi     = {10.1103/PhysRevE.99.062303}
}

@article{Mimar2021,
  author  = {Mimar, Sayat and Mussa Juane, Mariamo and Mira, Jorge and Park, Juyong and Mu{\~n}uzuri, Alberto P. and Ghoshal, Gourab},
  title   = {Linguistic evolution driven by network heterogeneity and the {Turing} mechanism},
  journal = {Phys. Rev. Research},
  volume  = {3},
  pages   = {023241},
  year    = {2021},
  doi     = {10.1103/PhysRevResearch.3.023241}
}

@article{Ghoshal2016,
  author  = {Ghoshal, Gourab and Mu{\~n}uzuri, Alberto P. and P{\'e}rez-Mercader, Juan},
  title   = {Emergence of a super-synchronized mobbing state in a large population of coupled chemical oscillators},
  journal = {Sci. Rep.},
  volume  = {6},
  pages   = {19186},
  year    = {2016},
  doi     = {10.1038/srep19186}
}

@article{GarciaSelfa2021,
  author  = {Garc{\'i}a-Selfa, David and Ghoshal, Gourab and Bick, Christian and P{\'e}rez-Mercader, Juan and Mu{\~n}uzuri, Alberto P.},
  title   = {Chemical oscillators synchronized via an active oscillating medium: Dynamics and phase approximation model},
  journal = {Chaos, Solitons \& Fractals},
  volume  = {145},
  pages   = {110809},
  year    = {2021},
  doi     = {10.1016/j.chaos.2021.110809}
}

@article{Stich2013,
  author  = {Stich, Michael and Ghoshal, Gourab and P{\'e}rez-Mercader, Juan},
  title   = {Parametric Pattern Selection in a Reaction-Diffusion Model},
  journal = {PLoS ONE},
  volume  = {8},
  number  = {10},
  pages   = {e77337},
  year    = {2013},
  doi     = {10.1371/journal.pone.0077337}
}

@article{Sowinski2024,
  author  = {Sowinski, Damian R. and Frank, Adam and Ghoshal, Gourab},
  title   = {Information-theoretic description of a feedback-control {Kuramoto} model},
  journal = {Phys. Rev. Research},
  volume  = {6},
  pages   = {043188},
  year    = {2024},
  doi     = {10.1103/PhysRevResearch.6.043188}
}
